# Structural stability of $Fe_5Si_3$ and $Ni_2Si$ studied by high-pressure x-ray diffraction and *ab initio* total-energy calculations

D. Errandonea[1,†,*] D. Santamaría-Perez[2,3], A. Vegas[3], J. Nuss[4], M. Jansen[4],

P. Rodríguez-Hernandez[5,*], and A. Muñoz[5,*]

[1]Departamento de Física Aplicada-ICMUV, Universitat de València, Edificio de Investigación, C/Dr. Moliner 50, 46100 Burjassot (Valencia), Spain

[2]Max Planck Institute fur Chemie, Postfach 3060, D-55020 Mainz, Germany

[3]Instituto de Química-Física Rocasolano, CSIC, C/Serrano 119, 28006 Madrid, Spain

[4]Max-Planck Institut für Festkörperforschung, Heisenbergstrasse 1, 70569 Stuttgart, Germany

[5]Departamento de Física Fundamental II, Universidad de La Laguna, La Laguna, Tenerife, Spain

**Abstract:** We performed high-pressure angle dispersive x-ray diffraction measurements on $Fe_5Si_3$ and $Ni_2Si$ up to 75 GPa. Both materials were synthesized in bulk quantities via a solid-state reaction. In the pressure range covered by the experiments, no evidence of the occurrence of phase transitions was observed. On top of that, $Fe_5Si_3$ was found to compress isotropically, whereas an anisotropic compression was observed in $Ni_2Si$. The linear incompressibility of $Ni_2Si$ along the *c*-axis is similar in magnitude to the linear incompressibility of diamond. This fact is related to the higher valence-electron charge density of $Ni_2Si$ along the *c*-axis. The observed anisotropic compression of $Ni_2Si$ is also related to the layered structure of $Ni_2Si$ where hexagonal layers of $Ni^{2+}$ cations alternate with graphite-like layers formed by $(NiSi)^{2-}$ entities. The experimental results are supported by *ab initio* total-energy calculations carried out using density functional theory and the pseudopotential method. For $Fe_5Si_3$, the calculations also predicted a phase transition at 283 GPa from the hexagonal $P6_3/mcm$ phase to the cubic structure adopted by Fe and Si in the garnet $Fe_5Si_3O_{12}$. The room-temperature equations of state for $Fe_5Si_3$ and $Ni_2Si$ are also reported and a possible correlation between the bulk modulus of iron silicides and the coordination number of their minority element is discussed. Finally, we report novel descriptions of these structures, in particular of the predicted high-pressure phase of $Fe_5Si_3$ (the cation subarray in the garnet $Fe_5Si_3O_{12}$), which can be derived from spinel $Fe_2SiO_4$ ($Fe_6Si_3O_{12}$).

† Author to whom correspondence should be addressed; FAX: (34) 96 354 3146; email: daniel.errandonea@uv.es
* Currently member of the MALTA Consolider Team

## I. Introduction

The Earth's core is believed to consist of an iron-nickel alloy with several percent of light alloying elements. In particular, silicon has been suggested, from geochemical arguments, as a possible major alloying element in the Earth's outer core. On top of that, iron silicides are also considered as probable candidates of the origin of the ultra-low velocity zone at the base of the Earth's mantle. These silicides may appear at the core-mantle boundary since liquid iron coexists with solid silicates. Because of these facts, the high-pressure structural stability of iron and nickel silicides is currently of interest to Earth scientists. Therefore, in order to understand the above described long-standing geophysical and geochemical subjects, a number of studies of iron silicides under pressure have been performed [1–7]. On the other hand, from a technological point of view, the ecologically friendly [7] iron and nickel silicides have also received a lot of attention since they have interesting magnetic [8, 9] and electronic properties [10, 11], which make them promising candidates for optoelectronic applications [10] and for the developing of metal-semiconductor contacts (Schottky junctions) [12]. Finally, iron and nickel silicides are known to produce oxides in which the Fe:Si (Ni:Si) stoichiometry is maintained. Thus, the oxidation of $Fe_5Si_3$ produces the garnet $Fe_5Si_3O_{12}$ whose $Fe_5Si_3$ subarray differs from the crystal structure of the silicide itself. In the same way, the high-temperature (HT) phase of $Ni_2Si$ ($\theta$-$Ni_2Si$) is isostructural to the $Ni_2Si$ subarray in the olivine-like $Ni_2SiO_4$. As in some cases, the cation compound of the produced oxides reproduces the structure of the high-pressure phase of the precursor materials [13 –16], it has been predicted that upon compression $Fe_5Si_3$ is expected to undergo a phase transition to a more compact structure, which may reproduce that of the cation array in the garnet $Fe_5Si_3O_{12}$. Also the room temperature (RT) phase of $Ni_2Si$ ($\delta$-$Ni_2Si$) could undergo a phase transition to the $MgCu_2$-type structure, in the same way that $Ni_2SiO_4$ undergoes the olivine-to-spinel phase transition. Because of the above given reasons, iron and nickel silicides are interesting compounds from geophysical, technological, and crystallochemical points of view.

In this work, we present new studies of the high-pressure structural stability of the mineral xifengite ($Fe_5Si_3$) and the first high-pressure structural studies of dinickel silicide ($Ni_2Si$), the main component of meteoritic nickel silicide (perryite). Room temperature angle dispersive x-ray diffraction (ADXRD) experiments were carried out in both silicides up to 75 GPa using a diamond-anvil cell (DAC). From our powder-diffraction experiments we concluded that $Fe_5Si_3$ remains stable in the hexagonal $P6_3/mcm$ low-pressure phase (space group (S.G.) No. 193) [3] and $Ni_2Si$ remains stable in the orthorhombic *Pbnm* structure (S.G. No. 62) - δ-$Ni_2Si$ phase - up to the highest pressure reached in our measurements. Both structures can be seen in figures 1 and 2. The experimental results were complemented by *ab initio* total-energy calculations that we performed using density functional theory (DFT) and the pseudopotential method. The theoretical results not only support the experimental results, but also predict the presence beyond 283 GPa of a new denser phase in $Fe_5Si_3$, which has the structure adopted by the $Fe_5Si_3$ subnet in the garnet $Fe_5Si_3O_{12}$. Beyond this, both theory and experiment found that the *c*-axis of δ-$Ni_2Si$ is much less compressible that its other two crystallographic axes. This fact is related to the bonding features of δ-$Ni_2Si$. Finally, accurate RT equation of states (EOS) for $Fe_5Si_3$ and δ–$Ni_2Si$ were obtained from the experimental data and the theoretical calculations. The reported results could have important implications for the differentiation processes of the planets and the composition of their cores.

## II. Experimental details

The synthesis of $Fe_5Si_3$ and $Ni_2Si$ was performed by a solid-state reaction from stoichiometric amounts of high-purity elements. The mixture was sealed under argon in a tantalum ampoule, annealed for two days at 1223 K (well below the eutectic temperature of the system [17]), and then quenched by exposure to air at RT. Gray metallic dendrites were obtained. Samples were prepared as finely ground powders from the synthesized materials immediately before the loading of the DAC to minimize any possible oxidation of the silicides. The silicide dendrites were ground in a marble mortar, which contained acetone (99.99% purity), using a marble pestle. The synthesized samples were characterized by x-ray diffraction at ambient conditions. There was no indication of additional phases in the starting materials. The unit-cell parameters were $a$ = 6.752(5) Å and $c$ = 4.741(3) Å for hexagonal $Fe_5Si_3$ and $a$ = 7.061(6) Å, $b$ = 4.992(4) Å, and $c$ = 3.741(2) Å for orthorhombic δ-$Ni_2Si$, which are in excellent agreement with previous studies [3, 17]. Several attempts to synthesize $Fe_2Si$ were also carried out, but we were not able to obtain a pure phase. High-pressure ADXRD measurements were carried out at RT in a 300-μm culet DAC. The powder samples were loaded together with a ruby chip into a 100-μm-diameter hole drilled on a 200-μm-thick rhenium gasket preindented to 35 μm. Silicone oil was used as pressure-transmitting medium [18, 19] and the pressure was determined using the ruby fluorescence technique [20]. The ADXRD experiments were performed at the 16-IDB beamline of the HPCAT facility at the Advanced Photon Source (APS) using monochromatic radiation with λ = 0.3931 Å. The monochromatic x-ray beam was focused down to 15 μm x 10 μm using Kickpatrick-Baez mirrors and spatially collimated with a 30 μm molybdenum clean-up pinhole. Diffraction images were recorded with a Mar345 image plate detector, located 350 mm away from the sample, and were integrated and corrected for distortions using the FIT2D software [21]. A $CeO_2$ standard was used to calibrate the detector parameters. Typical diffraction patterns were collected with 20 seconds exposures. Two independent runs were performed for each of the studied silicides. The indexing, structure solution, and refinements were performed using the DICVOL [22] and POWDERCELL [23] program packages.

**III. Overview of the calculations**

The structural stability of the phases of $Fe_5Si_3$, $Fe_2Si$ and $Ni_2Si$ was further investigated theoretically by means of total-energy calculations performed within the framework of density functional theory (DFT) with the Vienna *ab initio* simulation package (VASP) [24]. A review of DFT-based total-energy methods as applied to the theoretical study of phase stability can be found in Ref. [25]. In the calculations, the exchange and correlation energy was described within the generalized gradient approximation (GGA) described in Ref. [26]. We used ultrasoft pseudo-potentials and we adopted the projector augmented wave (PAW) scheme. We employed a basis set of plane waves up to a kinetic energy cutoff of 334.9 eV for $Fe_5Si_3$ and $Fe_2Si$, and 336.9 eV for $Ni_2Si$, and Monkhorts-Pack grids for the Brillouin-zone integrations which ensure highly converged and precise results [to about 1 meV per formula unit (pfu)]. At each selected volume for a given structure of the considered compound, the external and internal parameters were relaxed through the calculations of the forces on the atoms and the components of the stress tensor, which yielded the values of the atomic positions and unit-cell parameters of the structure. Valuable structural information (equilibrium volume, bulk modulus, etc.) for each stable phase was obtained from the calculated energy-volume curves after a Birch-Murnaghan fitting.

## IV. Results and discussion

### A. Structural studies of $Fe_5Si_3$

Figure 3 shows our ADXRD data for $Fe_5Si_3$ at several selected pressures and compares them with a diffraction pattern measured at atmospheric pressure (0.0001 GPa) outside the DAC. At ambient conditions, the obtained diffraction pattern corresponded to the hexagonal $Mn_5Si_3$-type structure (S.G. *P6₃/mcm*, No. 193) [3], with no indication of any additional phase in it. Under compression, the only changes we observed in the x-ray diffraction patterns are the typical peak broadening of DAC experiments [27, 28] and the appearance of a peak around $2\theta = 11^o$, denoted by the symbol * in figure 3. This peak has been assigned to a rhenium gasket line and can be easily identified since its pressure shift is smaller than that of the $Fe_5Si_3$ peaks. It is important to mention here that the gasket peak does not contaminate the x-ray diffraction pattern of $Fe_5Si_3$ since it does not overlap with any of the sample peaks. Regarding the $Fe_5Si_3$ peaks, we observed that they shift smoothly with compression and that all the Bragg reflections, present in the x-ray diffraction patterns, can be indexed within the $P6_3/mcm$ structure up to 75 GPa. The small changes of the relative intensities of some of the peaks can be assigned to preferred-orientation effects induced upon compression in the DAC [29, 30]. From the x-ray diffraction data, we obtained the evolution with pressure of the volume and lattice parameters of $Fe_5Si_3$. We also refined the atomic positions of the Fe and Si atoms. We found that, within the pressure range of our experiments, the pressure change of the *x* coordinate of the Fe and Si atoms, located at the Wyckoff position 6g (the only two free coordinates in the structure), is smaller than the experimental uncertainty. The mean values of these coordinates are $x_{Fe}$ = 0.230(4) and $x_{Si}$ = 0.599(4). Therefore, we concluded that the pressure effect on the atomic positions can be neglected, which is in good agreement with previous studies performed up to 30 GPa [3]. The pressure evolution of the unit-cell parameters of $Fe_5Si_3$ is plotted in figure 4, where we compare them with previously reported data obtained using NaCl as pressure medium [3] and with our theoretical calculations. Both experiments agree within its accuracy up to 20 GPa. Beyond this pressure, the previous experiment slightly underestimates the decrease of the volume. As it has been argued in the literature [31], this fact can be attributed to the larger non-hydrostatic stresses caused by the NaCl pressure medium used in the previous experiments [3].

In figure 4, it can be seen that the contraction of the unit-cell parameters with pressure is rather isotropic. Indeed, according with our experiments, the *c/a* ratio stays nearly equal to 0.702 within the covered pressure range. The same behavior was previously observed in Ref. [3] up to 30 GPa and our experiments verify that the compression of $Fe_5Si_3$ remains isotropic up to 75 GPa. A quadratic fit to our data gives the following pressure dependence of the unit-cell parameters of $Fe_5Si_3$:

$$a = 6.76(1) - 9.5(6)\ 10^{-3}\ P + 3.5(7)\ 10^{-5}\ P^2 \qquad \text{and}$$

$$c = 4.736(9) - 7.5(5)\ 10^{-3}\ P + 3.8(6)\ 10^{-5}\ P^2$$

where *a* and *c* are given in Å and P is in GPa. From these two relations, it can be estimated that from atmospheric pressure to 75 GPa *a* and *c* are reduced approximately a 7.5%. The isotropic compression of the unit-cell parameters and the fact that the atomic positions of Fe and Si do not change upon compression suggest that the only effect of pressure in the structure of $Fe_5Si_3$ is to produce a uniform change in all the bond distances as previously observed in ε-FeSi [32]. The evolution of the bond distances of $Fe_5Si_3$ with pressure has been calculated from our experimental data, being represented in figure 5. There, it can be seen that all the Fe-Fe and Si-Fe bonds follows a similar behavior upon compression. Additional support to this conclusion comes from the comparison of the bond distances reported at 0.0001 GPa and 30 GPa in Ref. [3].

The present pressure-volume data shown in figure 4 have been analyzed using a third-order Birch-Murnaghan EOS [33]. By fixing the zero-pressure volume ($V_0$) to its measured value (187.154 $Å^3$) we obtained the bulk modulus ($B_0$ = 215 ± 14 GPa) and its pressure derivative ($B_0$' = 3.6 ± 0.6). The bulk modulus obtained from our data is 13 % smaller than the value reported in Ref. [3] ($B_0$ = 243 ± 9 GPa), but our $B_0$' agrees within the uncertainties with the value reported in Ref. [3] ($B_0$' = 3.4 ± 0.9). The difference found for the bulk modulus, which is similar to the differences observed in the literature for other iron silicides [2, 32, 34, 35], may be caused by two reasons: 1) The data reported in Ref. [3] gives a smaller compressibility than the present data for $P \geq 20$ GPa. 2) The Murnaghan EOS [34] was used in Ref. [3] to fit $B_0$ and $B_0$' and this approach usually cause an overestimation of the bulk modulus. It is important to mention here, that after a comparison of the bulk modulus of $Fe_5Si_3$ with the bulk modulus of other iron silicides reported in the literature (e.g. the different polymorphs of fersilicite, FeSi, and ferdisilicite, $FeSi_2$), $Fe_5Si_3$ results to be the least compressible alloy. This can be seen in table 1, which summarizes the bulk modulus of different iron silicides. The compressibility of iron silicides has been previously proposed to be correlated with the coordination number (CN) of the minority element [3]. In the case of $Fe_5Si_3$ the silicon atoms have a CN = 9, whereas in the other alloys $6 \leq CN \leq 8$. This fact makes $Fe_5Si_3$ the least compressible compound among the different iron silicides studied up to now. According to this hypothesis, the hypothetical high-pressure phase of $Fe_5Si_3$ (the garnet-like), which has a CN = 10, and $Fe_2Si$, which has a CN = 11, should have a bulk modulus larger than 240 GPa. As we will show in the following, this is exactly what we have obtained from our *ab initio* calculations.

We compare now the experimental data presented with the results obtained from our total-energy calculations. Figure 6 shows the energy-volume curves for the different structures considered for $Fe_5Si_3$, from which the relative stability of the different phases can be extracted. Based upon either crystallochemical arguments or in its present observation in compounds related to $Fe_5Si_3$ we have considered the following structures in our calculations: hexagonal $Mn_5Si_3$-type (S.G. *P6₃/mcm*, No. 193) [3], tetragonal $Eu_5Si_3$-type (S.G. *I4/mcm*, No. 140) [42], orthorhombic $Sr_5Sb_3$-type (S.G. *Pnma*, No. 62) [43], and cubic $Ia\bar{3}d$ (S.G. No. 230) [3]. For the sake of clarity only the most competitive structures are shown in figure 6. This figure shows the hexagonal $P6_3/mcm$ structure to be stable up to 283 GPa, which agrees with the absence of phase transitions observed in the experiments up to 75 GPa. In addition, from the calculations we obtained the following EOS parameters for the $P6_3/mcm$ structure: $V_0$ = 175.7 Å$^3$, $B_0$ = 238.76 GPa and $B_0$' = 3.8 GPa. These values compares well with the experimental results, with differences within the typical reported systematic errors in DFT calculations. A similar degree of agreement exists for the calculated values of the internal parameters $x_{Fe}$ = 0.2450 for the Fe (6g) atoms, and $x_{Si}$ = 0.6044 for the Si (6g) atoms (experimental: 0.230 and 0.599, respectively) and *c/a* ratio = 0.719 (experimental: 0.702). The differences between the calculated volume and axial ratio can be mainly caused for an underestimation of the lattice parameter *a* by a 3 % (see figure 4). However the calculated pressure evolution of *a* follows a very similar trend than the experimental results. Our calculations also confirm that the compression of $Fe_5Si_3$ is isotropic up to 283 GPa; i.e. in the whole range of stability of the hexagonal phase of $Fe_5Si_3$. The calculations also found that there is no important effect of the pressure on the atomic positions of Fe and Si, in good agreement with our experiments.

As pressure increases, according with our calculations, the hexagonal $P6_3/mcm$ structure becomes unstable and converts into a body-centered cubic phase (S.G. $Ia\bar{3}d$, No. 230) with $Fe_1$ and $Fe_2$ atoms at 24c and 16a positions, respectively, and the Si atoms at the 24d sites; see figure 7. This high-pressure phase is isomorphous to the structure adopted by the $Fe_5Si_3$ subarray in the $Fe_5Si_3O_{12}$ garnet. This fact is in fully agreement with the hypothesis that proposes the existence of a correlation between oxidation and pressure [14, 15]. The high-pressure phase only emerges as thermodynamically stable above a compression threshold of about 283 GPa. From the common tangent construction or the enthalpy versus pressure plot [25], our calculations predict that $Fe_5Si_3$ becomes unstable in the *P6₃/mcm* phase at 283 GPa against the cubic $Ia\overline{3}d$ phase. The transition is a first-order phase transition with a volume change of 1.1 % and implies an increase of the Si coordination. The Si atoms (minority element) are coordinated by 9 Fe atoms in the low-pressure phase and by 10 Fe atoms in the high-pressure phase. The EOS fitting to the theoretical results gives $V_0$ = 346.60 Å$^3$, $B_0$ = 249.95 GPa, and $B_0'$ = 3.77 for the predicted high-pressure phase. This EOS indeed confirms that an increase of the coordination number of the minority element of the alloy should imply an increase of the bulk modulus.

In order to further check this hypothesis, we have also performed *ab initio* total-energy calculations for $Fe_2Si$. For this alloy two different polymorphs have been reported in the literature, the cubic hapkeite structure (S.G. $Pm\overline{3}m$, No. 221) [44], found in grains of lunar meteorites, and a trigonal structure (S.G. $P\overline{3}m1$, No. 164) [45], which is a slight distortion of the $Ni_2Al$-type structure. According with our calculations at zero and low pressure the most stable structure for $Fe_2Si$ is the trigonal structure reported by Kudielka [45], being the cubic hapkeite structure higher in energy by more than 200 meV pfu. Regarding the trigonal structure of $Fe_2Si$, we also found that in this structure, which we will name α-$Fe_2Si$, the three crystallographycally independent Fe atoms are located at the 1a (0, 0, 0); 1b (0, 0, ½) and 2d (⅓, ⅔, 0.73) sites and the Si atoms at 2d (⅓, ⅔, 0.25) sites, in good agreement with the experimental data [45]; see table 2. It is important to note that in α-$Fe_2Si$, the Si atoms (the minority element) are coordinated by 11 Fe atoms. The EOS fit to our theoretical results for α-$Fe_2Si$ gives $V_0$ = 64.985 Å$^3$, $B_0$ = 255 GPa, and $B_0'$ = 3.8. The obtained axial ratio for this structure is *c/a* = 1.245. The calculated ambient pressure volume underestimates the measured value of 72.3 Å$^3$, but the difference is within the typical systematic errors of DFT calculations. The calculated axial ratio is in very good agreement with the experimental value 1.255. Regarding the compressibility of α-$Fe_2Si$, our calculations show that this silicide has a large bulk modulus, 2% larger than that of the high-pressure phase of $Fe_5Si_3$, which gives additional support to the idea that relates the bulk modulus with the coordination number of the minority element in iron silicides.

**B. Structural studies of δ-$Ni_2Si$**

At atmospheric pressure, the obtained diffraction pattern for $Ni_2Si$ corresponded to the orthorhombic *Pbnm* structure (δ-$Ni_2Si$), with no indication of any extra phase in it. Under compression, we observed that all the $Ni_2Si$ peaks shift smoothly with compression and that all of them can be assigned to the *Pbnm* structure up to 75 GPa. From our x-ray diffraction data, we obtained the evolution with pressure of the volume and lattice parameters. We also obtained the atomic positions, being the Ni atoms located at two different 4c sites of coordinates (0.063, 0.325, 0.25) and (0.203, 0.042, 0.75) respectively, and the Si atoms also at 4c with coordinates (0.386, 0.263, 0.25). These positions agree with those reported in the literature [17] and the effect of pressure on them is comparable with the uncertainty of the experiments. The pressure dependences of the lattice parameters and the volume of $Ni_2Si$ are plotted in figure 8. The present pressure-volume data have been analyzed using a third-order Birch-Murnaghan EOS [33]. By fixing $V_0$ to its measured value (131.049 Å$^3$) we obtained $B_0$ = 167 ± 5 GPa and $B_0$' = 4.5 ± 0.5. The obtained bulk modulus is very similar to that of ε-Fe ($B_0$ = 163.4 ± 7.9 GPa and $B_0$' = 5.38 ± 0.16), the stable phase of iron at Earth's core conditions [46]. $Ni_2Si$ is known to have at atmospheric pressure a density (ρ) of 7.2 g/cm$^3$ and a melting temperature ($T_M$) of 1600 K. These values are close to those of pure iron, ρ = 7.8 g/cm$^3$ and $T_M$ = 1810 K [47]. Based upon these facts and the common presence of $Ni_2Si$ in meteorites, it has been speculated that nickel disilicide could be present in the core of the Earth [48]. As we will show later, our calculations suggest that $Ni_2Si$ remains stable in the orthorhombic *Pbnm* structure at Earth's inner core pressures. According to the present results, even at such extreme pressures, the difference between the density of δ-$Ni_2Si$ and ε-Fe stays close to 10 %.

In figure 8, it can be seen that the contraction of the unit-cell parameters with pressure is highly anisotropic. In particular the *c*-axis is much less compressible that the other two crystalline axes. A quadratic fit to our data reported in figure 8 gives the following pressure dependence of the unit-cell parameters of $Ni_2Si$:

$$a = 7.00(5) - 2.6(6)\ 10^{-2}\ P + 2.0(3)\ 10^{-4}\ P^2\ ,$$

$$b = 5.085(9) - 1.0(3)\ 10^{-2}\ P + 4(1)\ 10^{-5}\ P^2\ , \qquad \text{and}$$

$$c = 3.278(6) - 2.4(4)\ 10^{-3}\ P + 3.6(9)\ 10^{-6}\ P^2$$

where *a*, *b*, and *c* are given in Å and P is given in GPa. From these three relations it can be estimated that from atmospheric pressure to 75 GPa, *a* is reduced a 12.7%, *b* is reduced a 9.3%, and *c* is reduced a 3.9%. From these results, it can also be deduced that the linear incompressibility of δ-$Ni_2Si$ along the *c*-axis is 1450 GPa; i.e. it is similar to the linear incompressibility of diamond. This unique mechanical property would make δ-$Ni_2Si$ suitable for technological applications under extreme conditions. A better understanding of the observed anisotropic compressibility can be obtained from the analysis of the pressure evolution of the interatomic bond distances. We calculated the Ni-Ni and Ni-Si bond distances from our experimental data. The obtained results as a function of pressure are shown in figure 9. There, it can be seen that the atomic bonds that are mainly oriented perpendicular to the *c*-axis (solid symbols) are much more compressible that the atomic bonds oriented along the *c*-axis (empty symbols). This fact can be related to the anisotropic valence-electron density of $Ni_2Si$, mostly distributed along the *c*-axis [49]. This suggests that the directionality of the valence-electron density is the responsible of the large incompressibility of the *c*-axis of $Ni_2Si$.

This large incompressibility can also be understood by observing the structure represented in figure 2. When projected along the *c*-axis (figure 2a), one sees the classical description of δ-$Ni_2Si$ in terms of zigzag chains of $Ni_6Si$ trigonal prisms, further connected by edge-sharing to form blocks perpendicular to the *a*-axis [50]. However, in a recent re-interpretation of this structure-type [51], the δ-$Ni_2Si$ structure has been described as a Ni-stuffed, four-connected net, typical of the group 14 elements (see figure 2b). This structure is formed by puckered layers of hexagonal rings with some additional bonds between them. Within this framework, the structure of δ-$Ni_2Si$ can be thought as being formed by alternating hexagonal layers of $Ni^{2+}$ cations perpendicular to the *b*-axis and graphite-like layers perpendicular to the *b*-axis formed by $(NiSi)^{2-}$ entities (see figure 2c). This layered characteristic of δ-$Ni_2Si$ could be responsible of the observed difference of compressibility among different directions in $Ni_2Si$, as observed in other layered crystals like InSe [52] or $ReB_2$ [53].

We compare now the experimental data presented here with the results from our total-energy calculations. Figure 10 shows the energy-volume curves for most relevant structures among the several structures considered for $Ni_2Si$. From it, the relative stability of the different phases can be extracted. In order to theoretically test the structural stability of $Ni_2Si$, in addition to the δ-$Ni_2Si$ and θ-$Ni_2Si$ structures, we have selected some candidate structures adopted by other $A_2X$ compounds. These structures include the $Si_2Ti$-type structure (S.G. *Fddd*, No. 70) [54], the $Ni_2In$-type structure (S.G. $P6_3/mmc$, No. 194) [55], and the $MgCu_2$-type structure (S.G. $Fd\overline{3}m$, No. 227) [56], also know as the Cubic Laves Phase (C15 structure). Figure 10 shows the orthorhombic *Pbnm* structure to be stable up to nearly 400 GPa (the maximum pressure studied in our theoretical calculations), which agrees with the absence of phase transitions observed in the experiments up to 75 GPa. Our calculations also confirm that the compression of $Ni_2Si$ is highly anisotropic, being the *c*-axis the less compressible axis; see figure 8. In addition, the calculations give for the *Pbnm* structure of $Ni_2Si$ the following EOS parameters: $V_0$ = 133.44 Å$^3$, $B_0$ = 175.07 GPa, and $B_0$' = 5. The theoretically calculated EOS is shown together with the experimental data in figure 8. There, it can be seen that, in spite of the systematic volume overestimation, the *ab initio* calculations give a very similar compressibility than the experiments. The overestimation of the volume comes principally from the overestimation of the value of the unit-cell parameter *a*. The calculated values of the internal parameters agree also very well with the experimental values $x_{Ni1}$ = 0.0606, $y_{Ni1}$ = 0.3304, $x_{Ni2}$ = 0.2039, $y_{Ni2}$ = 0.0404, and $x_{Si}$ = 0.3855, $y_{Si}$ = 0.2862 for the 4c Ni and Si atoms.

## V. Crystal Chemistry of $Fe_5Si_3$, $Fe_2Si$ and $Ni_2Si$

### A. $Fe_2Si$, $Ni_2Si$ and the related oxides olivine and spinel $Fe_2SiO_4$

The only reference to the synthesis and structure elucidation of $Fe_2Si$ ($P\overline{3}m1$) was published by Kudielka [45]. As seen in figures 11(a-c), the reported structure is in between the $Ni_2Al$-type and the $Ni_2In$-type structures. From this figure it could be concluded that the $Ni_2Al$-type structure could transform into the $Ni_2In$-type by a continuous displacement of both $Fe_3$ and Si atoms. This fact could be related to the fact that theoretical calculations have some problems to determine which of the three structures (trigonal, $Ni_2Al$-type or $Ni_2In$-type) is the most stable in $Fe_2Si$. In particular, in the relaxation of these phases we found the existence of a number of local minima. These structurally different minima are located very close in energy, sometimes separated by shallow barriers, which make the precise determination of the absolute minimum within this set of crystal structures a rather tedious and difficult task. Indeed, theoretical calculations starting from the atomic coordinates reported for $Fe_2Si$ could convergence upon compression to the structural parameters of either, $Ni_2Al$-type or $Ni_2In$-type, being the former the most stable structure among these two. The lattice parameters of $Fe_2Si$ are given in table 2. When the *z* coordinates of both, $Fe_3$ and Si atoms, become ⅔ and 1/6, respectively, the $Ni_2Al$-type structure is produced. When fixed at ¾ and ¼, respectively, the $Ni_2In$-type ($P6_3/mmc$) structure is formed.

The important issue here is that the $Fe_2Si$ structure is very close to the $Ni_2In$-type which is the $Fe_2Si$ array existing in the olivine-like $Fe_2SiO_4$ (see figure 11d), as it was pointed out earlier [3]. That means that, when oxygen is inserted, the $Ni_2In$-type structure remains in the oxide, as in many other alloys [14, 15]. Thus, the following transitions can be observed either by inserting oxygen [14] or increasing pressure [57]:

$$Fe_2Si\ (P\bar{3}m1) \xrightarrow{O_2} Fe_2SiO_4\ \text{(olivine)} \xrightarrow{5\ \text{GPa}} Fe_2SiO_4\ \text{(spinel)}$$

The unsolved problem is, however, that the transformation $Ni_2In$-type-to-$MgCu_2$-type which occurs in the oxides, as the olivine-to-spinel transition, can not be predicted for the $Fe_2Si$ alloy in the pressure range covered by this work. This transition, which, as far as we know, has never been observed in alloys, should occur, for $Fe_2Si$ (also for $Ni_2Si$), at extremely high pressures.

As mentioned above, we tried the synthesis of $Fe_2Si$, but the impossibility of obtaining a pure phase, led us to substitute this compound by the related $Ni_2Si$. This compound, reported by Toman [17] is dimorphous. At RT it is of the $Co_2Si$-type ($\delta$-$Ni_2Si$) being strongly related to cotunnite, but at HT it transforms into the $Ni_2In$-type structure ($\theta$-$Ni_2Si$). Although the corresponding oxide $Ni_2SiO_4$ (olivine-like) undergoes the olivine-to-spinel transition, the corresponding transition in the alloy to the $MgCu_2$-type structure could not be observed up to 75 GPa. Theoretical calculations carried out in this work indicate that, even up to 400 GPa, this transition does not take place.

**B. $Fe_5Si_3$ and the garnet $Fe_5Si_3O_{12}$**

The three compounds studied here are related from a crystal chemical point of view. The structural behavior of the $Fe_5Si_3$ and its related oxide, the garnet $Fe_5Si_3O_{12}$, is summarized below:

$$Fe_5Si_3\ (P6_3/mcm)\ \text{xifengite} \xrightarrow{283\text{GPa}} Fe_5Si_3\ (Ia\bar{3}d) \qquad \text{[This work]}$$

$$Fe_5Si_3\ (P6_3/mcm) \xrightarrow{O_2} Fe_5Si_3O_{12}\ (Ia\bar{3}d)\ \text{skiagite garnet} \qquad \text{[Ref. 58]}$$

The structure of xifengite is represented in figure 1, projected on the *ab* plane. It belongs to the $Mn_5Si_3$-type and is also adopted by several cation arrays, like $Ca_5P_3$ in apatite ($Ca_5P_3O_{12}F$).

In this article, however, the description will focus on other aspects, such as the coordination polyhedron of the Fe atoms around the Si atoms, because the CN normally increases with pressure. As seen in table 1, this feature is helpful to rationalize the bulk modulus found for the different phases of iron silicides. In $Fe_5Si_3$, the Si atoms are surrounded by 9 Fe atoms forming a polyhedron which can be seen as a distorted square antiprism, with an additional Fe atom capping one of the square faces.

As discussed above, theoretical calculations predict that, at 283 GPa, the silicide xifengite transforms into a cubic structure which coincides with the cation array of the garnet $Fe_5Si_3O_{12}$ (skiagite) [58]. Because the garnet structure is rather complicated we see necessary a comprehensive description of this oxide. Both compounds, the alloy $Fe_5Si_3$ (P > 283 GPa) and the oxide $Fe_5Si_3O_{12}$, are cubic ($Ia\bar{3}d$, Z = 8) with unit-cell parameters $a$ = 7.02 Å and $a$ = 11.73 Å, respectively. Their atomic coordinates are given in table 3.

The classical description of the garnet structure can be found in text books devoted to structural chemistry [59]. This description emphasizes the cation-centered, oxygen coordination polyhedra. Thus, Si atoms form isolated tetrahedral orthosilicate groups ($SiO_4$). The Fe atoms occupy two crystallographycally independent positions and have two different coordination polyhedra. $Fe_2$ are at the center of O octahedra and the $Fe_1$ atoms are surrounded by eight O atoms forming a rather distorted cube. Its crystal chemical notation would be $Fe^{[8]}{}_3Fe^{[6]}{}_2Si^{[4]}{}_3O_{12}$. This simple description was considered unsatisfactory by O'Keeffe and Hyde [60] in their alternative approach describing crystal structures as oxygen-stuffed alloys.

These authors noticed that the cation array of the garnet-like structure was related to that of the $Cr_3Si$ alloy. Both structures are represented in figure 12. In part (a) we have drawn a complete unit cell ($a$ = 4.55 Å) of the $Cr_3Si$ structure ($Pm\bar{3}n$, Z = 2) [61]. It is formed by a body-centered cubic (bcc) array of Si atoms whose faces are centered by pairs of Cr atoms separated at short distances of 2.27 Å. The twelve Cr atoms, when connected, form an irregular icosahedron centered by the Si atom. In figure 12b it is represented 1/8 of the unit cell of the garnet $Fe_5Si_3O_{12}$ ($a$ = 11.73 Å). As seen, both structures are topologically identical. In the case of the oxide, the unit cell is doubled because both, $Fe_1$ and Si atoms alternate at the face centers. It must be outlined that $Ia\bar{3}d$ becomes a subgroup of the type IIa by doubling the unit cell of the S.G. $Pm\bar{3}n$. On the contrary, here, the bcc array is formed by the $Fe_2$ atoms. It can be concluded that the similarities between both compounds are merely topological but that the atomic species and hence, the superstructure formed in the oxide, must obey to significant chemical differences. In connection with this, it is worth remarking that the center of the icosahedrons are occupied in both compounds by an isolated atom (Si and $Fe_2$ respectively), in spite of being the biggest hole in the structure (CN 12). On the contrary, in skiagite, the faces are centered by $SiO_4$ groups, instead of the Cr atoms of the compound.

However, it has been outlined elsewhere [14, 15] that the alloys described in Ref. [59] are mostly non-existing and that an extension of their approach, based on real stuffed alloys, was further proposed by Vegas *et al.* [14, 15]. As said in the introduction, this study was undertaken in part as an additional proof of this approach. Our theoretical calculations predict the xifengite-to-garnet-like transition at high pressure. The structure of the high-pressure iron silicide is identical to the same array in the garnet. Thus, this is a new example of a *real* oxygen-stuffed alloy.

A question which arises from the above discussion is why the Si and Fe atoms interchange their role in both compounds. This problem which could not be solved by a simple topological comparison of both structures, $Cr_3Si$ and $Fe_5Si_3O_{12}$, could find a solution by comparing the garnet structure with other related iron silicates, such as $Fe_2SiO_4$.

It is well known that $Fe_2SiO_4$ (olivine-like at ambient pressure), transforms into the spinel structure at 5 GPa [57]. Following O'Keeffe and Hyde, [60] the $Fe_2Si$ array in spinels ($MgCu_2$-type) can be seen as a three-dimensional network of Fe tetrahedral arrays, sharing all corners. This array corresponds to one half of the atoms forming a face-centered-cubic structure and the missing atoms originating big voids which are truncated tetrahedrons formed by 12 Fe atoms, and where the Si atoms ($SiO_4$ groups) are located. One of these truncated tetrahedra is represented in figure 13a. In the high-pressure transition, the Si atoms increase their CN from 11 (in olivine) to 12 (in spinel). Another interesting aspect of the spinel structure is that the Si atoms alone form a diamond-like (really a Si-like) network in spite of being formed by isolated orthosilicate groups.

Once we have described the cation array in the spinel structure, we are in conditions of establishing a novel structural relationship with that of garnet. Their relation is better deduced if we put their formula on the same basis:

$Fe_6Si_3O_{12}$ (spinel) - $Fe_5Si_3O_{12}$ (garnet)

Looking at the formulae, we could convert the spinel structure into the garnet structure, by only eliminating 1/6 of the Fe atoms. The important question, here, is whether the elimination of one Fe atom can be considered as a mere *gedanken experiment* which makes the two compounds to have almost the same empirical formula or, on the contrary, this small change in composition only produces, proportionally, small changes in the structure. If this were so, then the garnet structure must preserve important similarities with the structure of spinel. The answer to this crucial question is that both structures are strongly related.

We discussed above that the Fe atoms, in spinel, form a 3D network of corner-connected tetrahedra. In the garnet structure, however, the Fe atoms form a 3D network of trigonal bipyramids (two tetrahedra with a common base) whose corners are all shared with adjacent bipyramids. Thus, as consequence of the lower Fe contents, the tetrahedra condense in denser groups. This condensation also produces, however, big voids which are not anymore the truncated tetrahedron represented in figure 13a, but a new polyhedron formed by 10 Fe atoms. As expected, the CN decreases from 12 to 10. Like in spinel, the $SiO_4$ groups are also located at the center of these voids. The differences are well illustrated in figure 13. In the part (a) was drawn the truncated tetrahedron (12 Fe atoms) of spinels and in part (c) we have represented the new polyhedron formed in garnets. The transition between them can be achieved by eliminating one of the two atoms involved in opposite edges of figure 13a. When these two atoms disappear, the two remaining Fe atoms migrate to the center of the edges producing so the polyhedron of the garnet. These new Fe (migrated) atoms have been drawn in red in the central drawing (figure 13b). Note that in figure 13b, the original Fe atoms (existing in spinel) have been maintained together with the new (red) Fe atoms (existing in garnet). In this way, the transformation between both structures is clarified. It should be added that the diamond-like skeleton formed by the Si atoms in the spinel structure remains in the garnet. This is an additional indication that both compounds are strongly related.

Finally, it should be outlined that a garnet-like structure has been obtained in the $MgSiO_3$ catena-silicate under pressure [62]. This fact is not surprising if we look at its stoichiometry ($Mg_3^{[8]}(Mg,Si)^{[6]}Si_3^{[4]}O_{12}$). Note that it is isoelectronic to both, $Fe_3Fe_2Si_3O_{12}$ and $Nd_3Al_2Al_3O_{12}$ compounds. In connection with this, it could be speculated whether a garnet-like structure could exist for $Al_2O_3$ ($Al_3^{[8]}Al_2^{[6]}Al_3^{[4]}O_{12}$).

## VI. Concluding Remarks

The high-pressure structural stability of $Fe_5Si_3$ and $Ni_2Si$ have been studied by means of x-ray diffraction experiments as well as by *ab initio* calculations. In our experiments we observed the absence of phase transitions in both, $Fe_5Si_3$ and δ-$Ni_2Si$ up to 75 GPa. We also found that the compression of $Fe_5Si_3$ is rather isotropic whereas the compression of $Ni_2Si$ is highly anisotropic. This anisotropic behavior seems to be correlated with the spatial orientation of the chemical bonds and the highly oriented valence-electron density of $Ni_2Si$. The experimental results are supported by the *ab* *initio* total-energy calculations, which also predict the occurrence of a phase transition in $Fe_5Si_3$ at 283 GPa from the hexagonal *$P6_3/mcm$* phase to a cubic phase belonging to space group $Ia\bar{3}d$ (structure of the $Fe_5Si_3$ cation subarray in the garnet $Fe_5Si_3O_{12}$). On the other hand, the *ab initio* calculations predict that the orthorhombic *Pbnm* structure of δ-$Ni_2Si$ is the most stable structure at least up to 400 GPa. Finally, an EOS was determined for $Fe_5Si_3$ ($Ni_2Si$) giving the following parameters: $V_0$ = 187.154 Å$^3$, $B_0$ = 215 ± 14 GPa, and $B_0$' = 3.6 ± 0.6 ($V_0$ = 131.049 Å$^3$, $B_0$ = 147 ± 5 GPa, and $B_0$' = 4.5 ± 0.5) for the low-pressure phase and $V_0$ = 346.60 Å$^3$, $B_0$ = 249.95 GPa, and $B_0'$ = 3.77 for the high-pressure phase of $Fe_5Si_3$. The prediction of the cubic phase ($Ia\bar{3}d$) for $Fe_5Si_3$, under high pressure, is in agreement with the concept that relates oxidation and pressure [14, 15]. The crystal chemistry of $Fe_5Si_3$, $Fe_2Si$, and $Ni_2Si$ is also systematically discussed. It is shown that the $Fe_2Si$ structure, as well as the cation array in the olivine-like $Fe_2SiO_4$, can be seen as a continuous displacive transition from the $Ni_2Al$-type structure. In the same way, a new structural relationship, with more physical meaning, can be established between the high-pressure phase of $Fe_2SiO_4$ (spinel) and the garnet $Fe_5Si_3O_{12}$.

**Acknowledgments**

This work was made possible through financial support of the MCYT of Spain under Grants No. MAT2004-05867-C03-01, MAT2004-05867-C03-02, MAT2004-05867-C03-03, MAT2007-65990-C03-01, and MAT2007-65990-C03-03. Three of the authors (D.E., P.R.-H., and A.M.) are also indebted to the MCYT of Spain for additionally supporting their own research with the grant No. CSD2007-00045. D.E. acknowledges the financial support from the MCYT of Spain through the "Ramón y Cajal" program. He also wants to thank the Alexander von Humboldt Foundation for its generous support. The U.S. Department of Energy, Office of Science, Office of Basic

Energy Sciences supported the use of the Advanced Photon Source (APS) under Contract No. W-31-109-Eng-38. DOE-BES, DOE-NNSA, NSF, DOD-TACOM, and the W.M. Keck Foundation supported the use of the HPCAT facility. We would like to thank S. Sinogeikin, Y. Meng, and the rest of the staff at the HPCAT of the APS for their contribution to the success of the ADXRD experiments. We also thank O. Tschauner and R. Kumar from UNLV for sharing part of their beam time with us.

## References

[1] D.P. Dobson, L. Vocadlo, and I.G. Wood, Am. Mineral. **87**, 784 (2002).

[2] J. F. Lin, A. J. Campbell, D.L. Heinz, and G.Y. Shen, J. Geophysical Research **108**, 2045 (2003).

[3] D. Santamaria-Perez, J. Nuss, J. Haines, M. Jansen, and A. Vegas, Solid State Sciences **6**, 673 (2004).

[4] Y. Kuwayama and K. Hirose, Am. Mineral. **89**, 273 (2004).

[5] K. Takarabe, T. Ikai, Y. Mori, H. Udono, and I. Kikuma, J. Appl. Phys. **96**, 4903 (2004).

[6] D. Santamaría-Pérez and R. Boehler, Earth Planet. Sci. Lett. **265**, 743 (2008).

[7] K. Tajima, Y. Endoh, J.E. Fischer, and G. Shirane, Phys. Rev. B **38**, 6954 (1988).

[8] D. Songlin, O. Tegus, E. Bruck, J.C.P. Klaase, F. R. De Boer, and K.H.J. Buschow, J. Alloys Compd. **334**, 249 (2002).

[9] F. Riedel and W. Schroter, Phys. Rev. **62**, 7150 (2000).

[10] D. Leong, M. Harry, K.J. Reeson, and K.P. Homewood, Nature **387**, 686 (1997).

[11] K. Takakura, T. Suemasu, Y. Ikura, and F. Hasegawa, J. Appl. Phys. Jpn. **39**, L789 (2000).

[12] H. Lange, Phys. Stat. Sol. B **101**, 3 (1997).

[13] L. A. Martínez-Cruz, A. Ramos-Gallardo, and A. Vegas, J. Solid State Chem. **110**, 397 (1994).

[14] A. Vegas, Crystallogr. Rev. **7**, 189 (2000).

[15] A. Vegas and M. Jansen, Acta Cryst. B **58**, 38 (2002).

[16] A. Vegas, A. Grzechnik, K. Syasen, I. Loa, M. Hanfland, and M. Jansen, Acta Cryst. B **57**, 151 (2001).

[17] K. Toman, Acta Cryst. **5**, 319 (1952).

[18] Y. R. Shen, R. S. Kumar, M. Pravica, and M. Nicol, Rev. Scientific Instr. **75**, 4450 (2004).

[19] D. Errandonea, Y. Meng, M. Somayazulu, and D. Häusermann, Physica B **355**, 116 (2005).

[20] H. K. Mao, J. Xu, and P. M. Bell, J. Geophys. Res. **91**, 4673 (1986).

[21] A. P. Hammersley, S. O. Svensson, M. Hanfland, A. N. Fitch, and D. Häusermann, High Press. Res. **14**, 235 (1996).

[22] A. Boultif and D. Louer, J. Appl. Crystal. **24**, 987 (1991).

[23] W. Kraus and G. Nolze, J. Appl. Crystal. **29**, 301 (1996).

[24] G. Kresse and J. Furthmüller, Comput. Mater. Sci. **6**, 15 (1996); Phys. Rev. B **54**, 11169 (1996).

[25] A. Mujica, A. Rubio, A. Muñoz, and R. J. Needs, Rev. Mod. Phys. **75**, 863 (2003).

[26] J. P. Perdew, K. Burke and M. Ernzerhof, Phys. Rev. Lett. **77,** 3865 (1996).

[27] D. He and T. Duffy, Phys. Rev. B **73**, 134106 (2006).

[28] D. Errandonea, R. Boehler, S. Japel, M. Mezouar, and L. R. Benedetti, Phys. Rev. B **73**, 092106 (2006).

[29] O. Tschauner, J. McClure, and M. Nicol, J. Synchrot. Radiat. **12**, 626 (2005).

[30] D. Errandonea, R. Boehler, B. Schwager, and M. Mezouar, Phys. Rev. B **75**, 014103 (2007).

[31] D. Errandonea, J. Pellicer-Porres, F. J. Manjón, A. Segura, Ch. Ferrer-Roca, R. S. Kumar, O. Tschauner, P. Rodriguez-Hernandez, S. Radescu, J. Lopez-Solano, A. Mujica, A. Muñoz, and G. Aquilanti, Phys. Rev. B **72**, 174106 (2005).

[32] I. G. Wood, W. I. F. David, S. Hull, and G. D. Price, J. Appl. Crystal. **29**, 215 (1996).

[33] F. Birch, J. Geophys. Res. **83**, 1257 (1978).

[34] L. Vocadlo, G.D. Price, and I.G: Wood, Acta Crystal. B **56**, 369 (2000).

[35] K. Takarabe, R. Teranishi, J. Oinuma, and Y. Mori, J. Phys. Condens. Matter **14**, 11007 (2002).

[36] F. D. Murnaghan, Proc. Nat. Acad. Sciences **30**, 244 (1949).

[37] D. P. Dobson, W. A. Crichton, P. Bouvier, L. Vocadlo, and I. G. Wood, Geophys. Res. Letters **30**, 1014 (2003).

[38] K. Takarabe, T. Ikai, Y. Mori, H. Udono, and I. Kikuma, J. Appl. Phys. **96**, 4903 (2004).

[39] N. Hirao, E. Ohtani, T. Kondo, and T. Kikegawa., Phys Chem. Minerals **31**, 329 (2004).

[40] E. G. Moroni, W. Wolf, J. Hafner, and R. Podloucky, Phys. Rev. B **59**, 12860 (1999).

[41] S. Ono, T. Kikegawa, and Y. Ohishi, Europ. Journal Mineral. **19**, 183 (2007).

[42] R. Pöttgen, R.-D. Hoffmann, and D. Kußmann, Z. Anorg. Allg. Chem. **624**, 945 (1998).

[43] M. Martinez-Ripoll and G. Brauer, Acta Cryst. B **29**, 2717 (1973).

[44] M. Chen, J. Shu, H. K. Mao, X. Xie, and R. J. Hemley, Proc. Natl. Acad. Sci. USA **100**, 14651 (2003).

[45] H. Kudielka, Z. Kristallogr. **145**, 177 (1977).

[46] A. Dewaele, P. Loubeyre, F. Occelli, M. Mezouar, P. I. Dorogokupets, and M. Torrent, Phys. Rev. Let. **97**, 215504 (2006).

[47] R. Boehler, Nature **363**, 534 (1993); D. Errandonea, B. Schwager, R. Ditz, C. Gessmann, R. Boehler, and M. Ross, Phys. Rev. B **63**, 132104 (2001).

[48] J. M. Herndon, Proc. Nat. Acad. Sci. USA **93**, 646 (1996).

[49] A. Franciosi, J. H. Weaver, and F. A. Schmidt, Phys. Rev. B **26**, 546 (1982).

[50] A. Vegas and M. Martínez-Ripoll, Acta Cryst. B **48**, 747 (1992).

[51] A. Vegas and V. García-Baonza, Acta Cryst. B **63,** 339 (2007).

[52] D. Errandonea, D. Martínez-García, A. Segura, A. Chevy, G. Tobias, E. Canadell, and P. Ordejón, Phys. Rev. B **73**, 235202 (2006).

[53] H. Y. Chung, M. B. Weinberger, J. B. Levine, A. Kavner, J. M. Yang, S. H. Tolbert, and R. B. Kaner, Science **316**, 436 (2007).

[54] F. Laves and H. J. Wallbaum, Z. Kristallogr. **101**, 78 (1939).

[55] F. Laves and H.J. Wallbaum, Z. Ang. Miner. **4**, 17 (1942).

[56] R. W. Wyckoff, *Crystal Structures* (John Wiley and Sons, New York, 1964).

[57] S. I. Akimoto, H. Fujisawa, and T. Katsura, J. Geophys. Res. **70**, 1969 (1965).

[58] G. Ottonello, M. Borketa, P.F. Sciuto, Amer. Mineral. **81**, 429 (1996).

[59] A. F. Wells, *Structural Inorganic Chemistry* (Clarendon Press, Oxford, 1975).

[60] M. O'Keeffe and B.G. Hyde, *Structure and Bonding* Vol. 61, pp. 77-144 (Springer Verlag, Berlin, 1985).

[61] J. E. Jorgensen and S. E. Rasmussen, Acta Cryst. B **38**, 346 (1982).

[62] R. J. Angel, L. W. Finger, R. M. Hazen, M. Kanzaki, D. J. Weidner, R. C. Liebermann, and D.R. Veblen, Amer. Mineral. **74**, 509 (1989).

**Table 1**: Bulk modulus and coordination number of the minority element in different iron silicides.

| Compound | $B_0$ (GPa) | CN of minority element | Reference |
|---|---|---|---|
| $\alpha$-$FeSi_2$ | 167 | 6 | [40] |
| $\alpha$-$FeSi_2$ | 172 –182 | 6 | [37, 40] |
| $Fe_3Si$ | 182 | 7 | [40] |
| $\varepsilon$-FeSi | 160 – 200 | 7 | [31, 33] |
| CsCl-FeSi | 184 – 225 | 8 | [36, 41] |
| $\beta$-$FeSi_2$ | 180 – 200 | 8 | [34, 39] |
| $Fe_7Si_3$ | 199 - 207 | 8 | [2, 38] |
| hex-$Fe_5Si_3$ | 215 ± 14 | 9 | This work experimental |
| hex-$Fe_5Si_3$ | 239 | 9 | This work calculated |
| hex-$Fe_5Si_3$ | 243 ± 9 | 9 | [3] |
| cubic-$Fe_5Si_3$ | 250 | 10 | This work calculated |
| $Fe_2Si$ | 255 | 11 | This work calculated |

**Table 2:** Structural parameters of $Fe_2Si$. Data taken from Ref. [45], space group: $P\bar{3}m1$, $a$ = 4.052 Å and $c$ = 5.085 Å.

| Atom | Site | x | y | z |
|---|---|---|---|---|
| $Fe_1$ | 1a | 0 | 0 | 0 |
| $Fe_2$ | 1b | 0 | 0 | 0.5 |
| $Fe_3$ | 2d | 1/3 | 2/3 | 0.78 |
| $Si_1$ | 2d | 1/3 | 2/3 | 0.28 |

**Table 3:** Atomic coordinates for the high-pressure phase of $Fe_5Si_3$ and the garnet $Fe_5Si_3O_{12}$. Both compounds are cubic (S.G. $Ia\bar{3}d$) with lattice parameters $a$ = 7.02 Å and $a$ = 11.73 Å, respectively.

| Atom | Site | x | y | z |
|---|---|---|---|---|
| O (garnet) | 96h | 0.03529 | 0.05288 | 0.65769 |
| $Fe_1$ | 24c | 0.125 | 0 | 0.25 |
| $Fe_2$ | 16a | 0 | 0 | 0 |
| Si | 24d | 0.375 | 0 | 0.25 |

## Figure Captions

**Figure 1:** The structure of $Fe_5Si_3$ projected along the *ab* plane. Dark (light) circles correspond to the Fe (Si) atoms.

**Figure 2:** The structure of $Ni_2Si$. (a) Projection along the *c*-axis showing the distorted trigonal prism of Ni around the Si atoms. (b) Schematic view to show the δ-$Ni_2Si$ as a Ni-stufed four coordinated net. (c) Projection along the *b*-axis showing the presence of $(NiSi)^{2-}$ graphite-like layers. Ni and Si atoms are identified in the figure.

**Figure 3:** Room-temperature ADXRD data of $Fe_5Si_3$ at different pressures. In all diagrams the background was subtracted. The symbols * indicate the position of a gasket peak.

**Figure 4:** Volume and lattice parameters of $Fe_5Si_3$ under pressure. The solid circles represent the present data. The empty circles correspond to data of Ref. [3]. The solid lines are the reported EOS and quadratic fits to *a* and *c*. The dashed line represents the EOS reported in Ref. [3], and the dotted lines the theoretical results.

**Figure 5:** Pressure dependence of the Fe-Fe and Si-Fe bond distances for the low-pressure phase of $Fe_5Si_3$.

**Figure 6:** Total-energy versus volume from *ab initio* calculations for the analysed structures of $Fe_5Si_3$. Only the most competitive structures are shown. (Volume and energy are per two formula units).

**Figure 7:** The structure of $Fe_2Si$ projected in the *ab* plane. Dark (light) circles correspond to the Fe (Si) atoms.

**Figure 8:** Volume and lattice parameters of $Ni_2Si$ under pressure. The solid circles represent the present data. The solid lines are the reported EOS and quadratic fits to *a*, *b*, and *c* and the dotted lines the theoretical results.

**Figure 9:** Pressure dependence of the Ni-Ni and Si-Ni bond distances for $Ni_2Si$. The empty (solid) symbols represent the bonds orientated along (perpendicular to) the *c*-axis. The lines are quadratic fits to the experimental data.

**Figure 10:** Total-energy versus volume from *ab initio* calculations for the analyzed structures of $Ni_2Si$. Only the most competitive structures are shown. (Volume and energy are per two formula units).

**Figure 11:** The structures of (a) $Fe_2Si$, (b) $Ni_2Al$, (c) $Ni_2In$, and (d) $Fe_2SiO_4$. Different atoms are identified in the figure.

**Figure 12:** The structure of $Cr_3Si$ (a) and the cation subarray of garnet $Fe_2SiO_4$ (b). Different atoms are identified in the figure.

**Figure 13 (color online)**: Relationship between the spinel (a) and garnet (b) polyhedra. The dark (light) circles represent the Fe (Si) atoms. The Fe atom migrating to the center of the edges to produce the garnet polyhedron are shown in red.

**Figure 1**

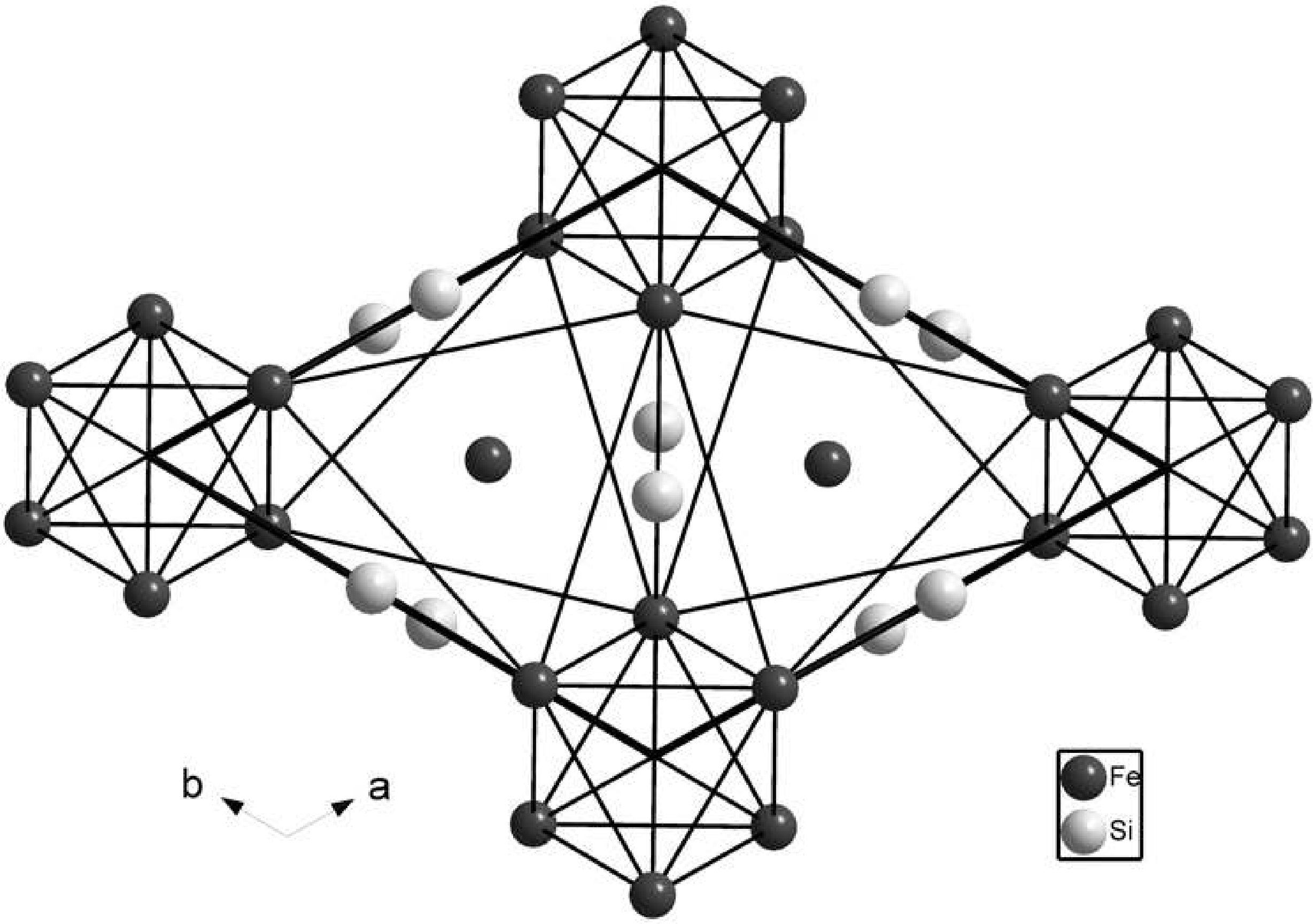

**Figure 2**

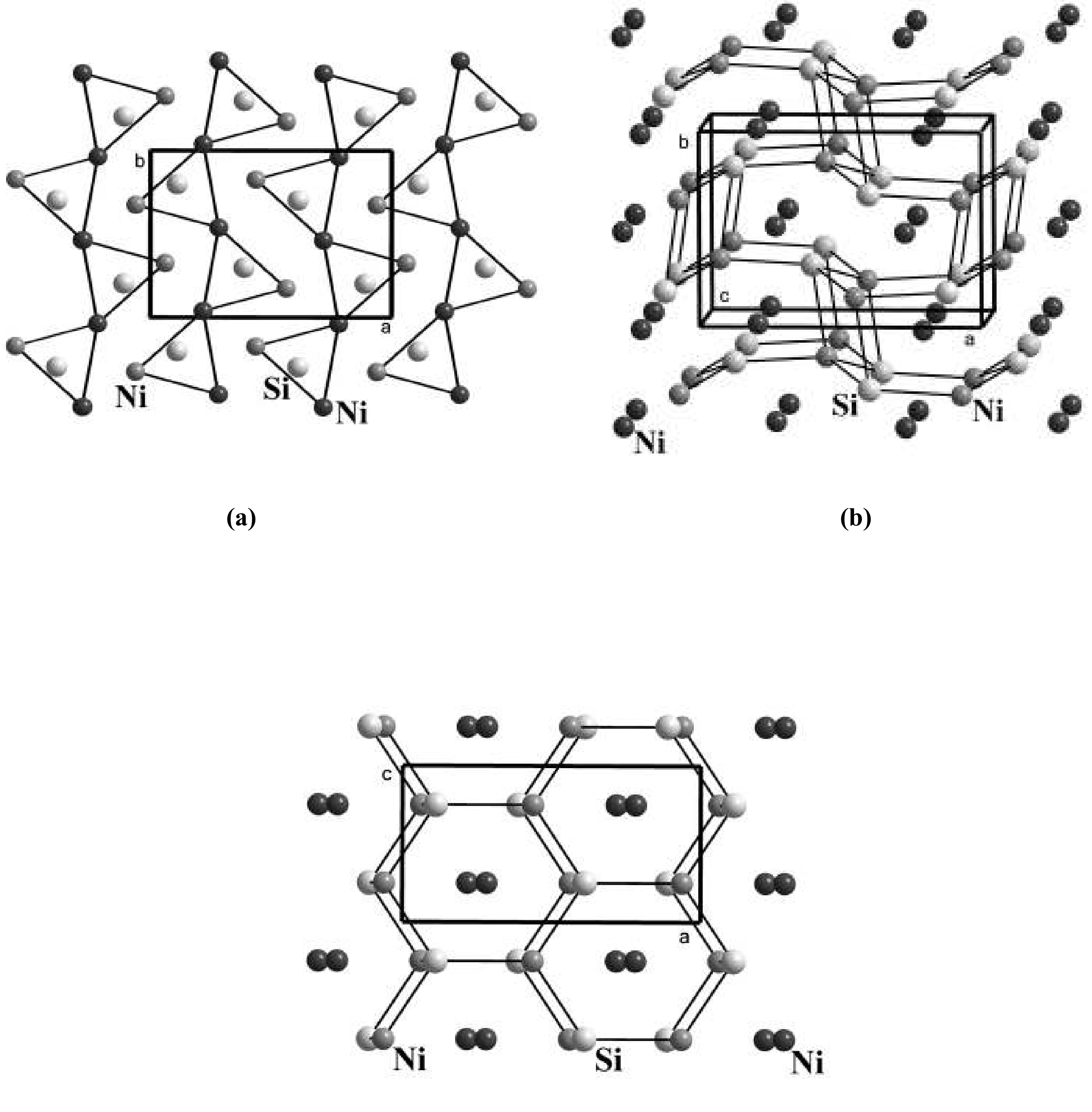

(a) (b) (c)

**Figure 3**

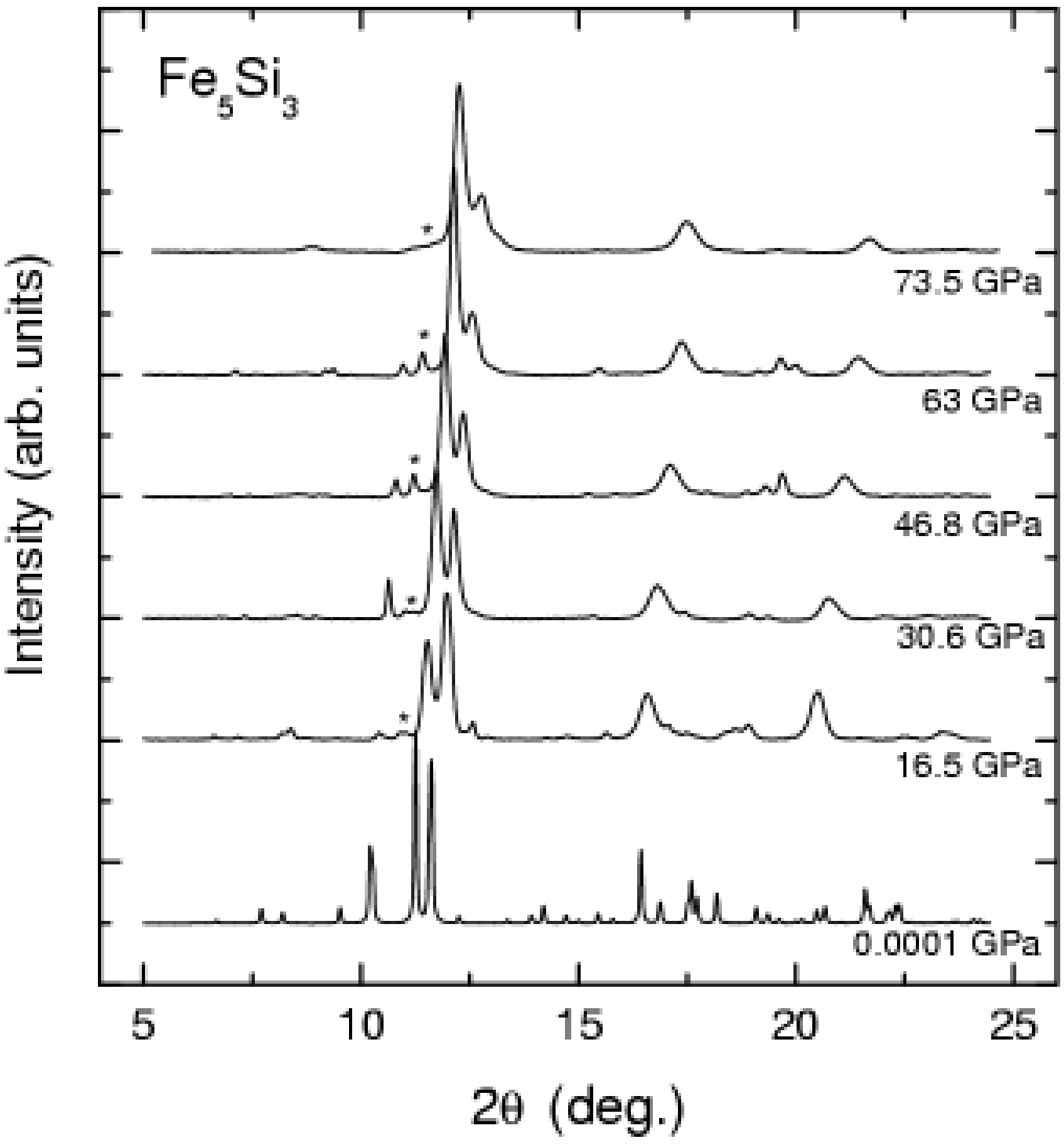

**Figure 4**

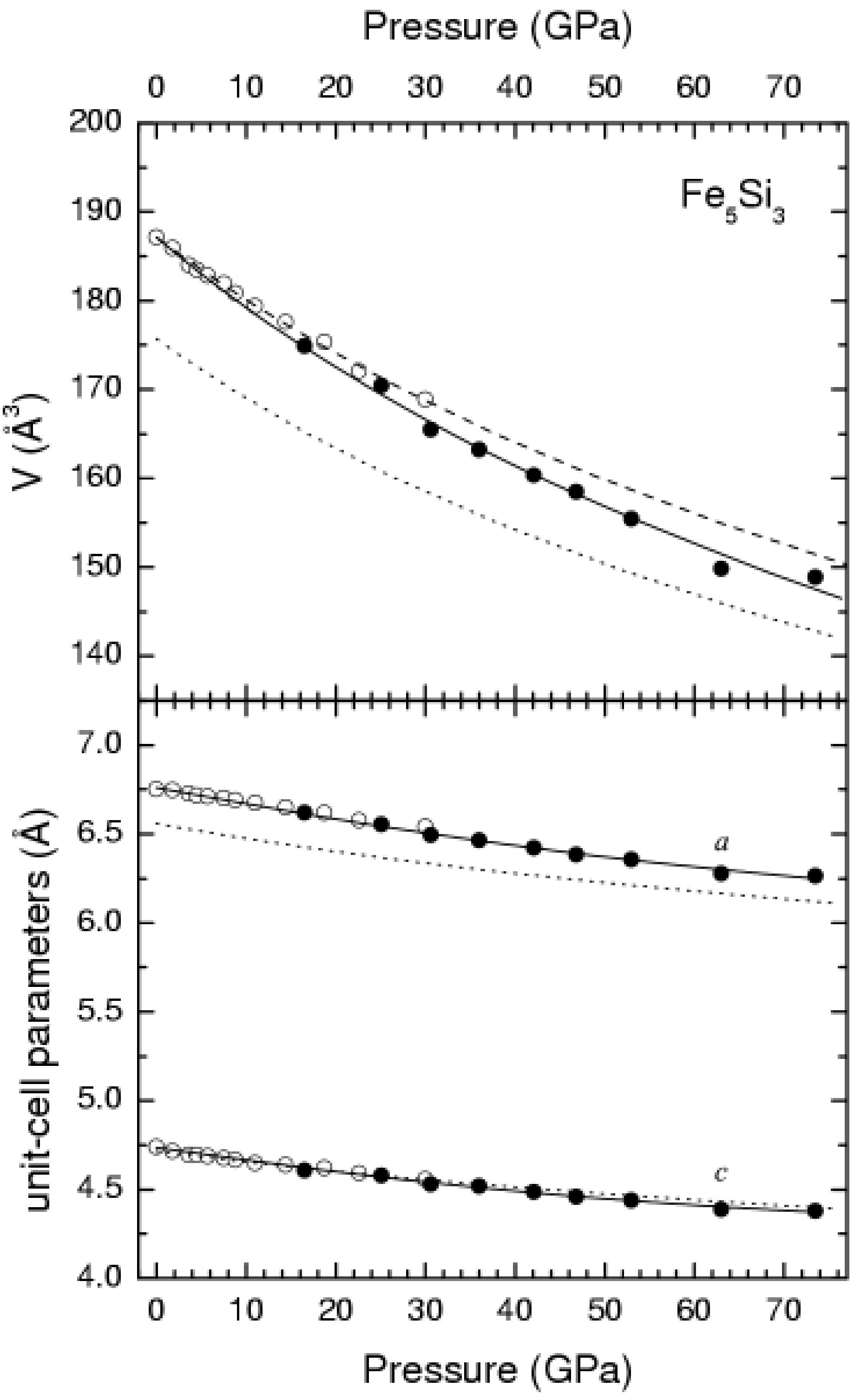

**Figure 5**

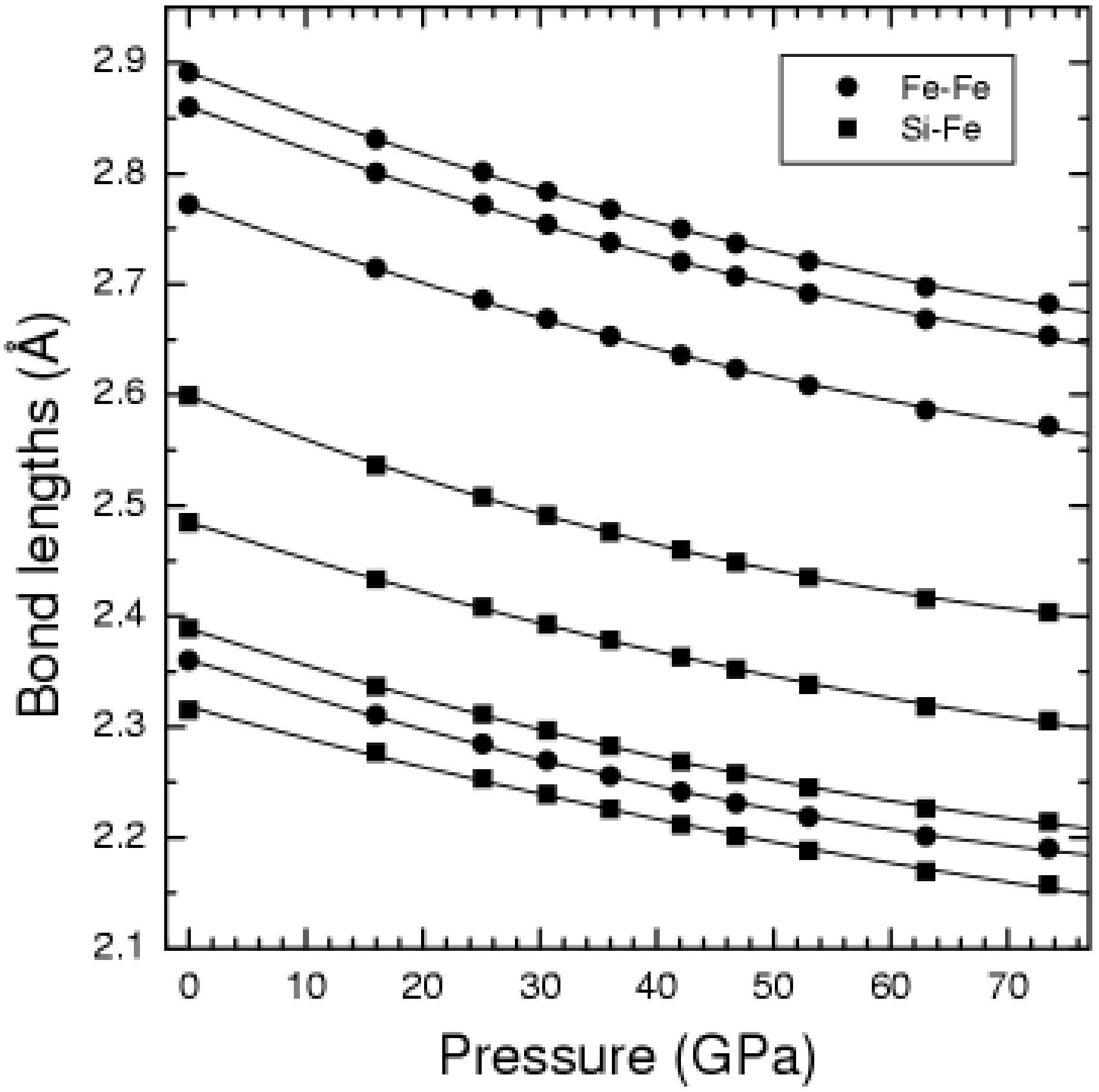

**Figure 6**

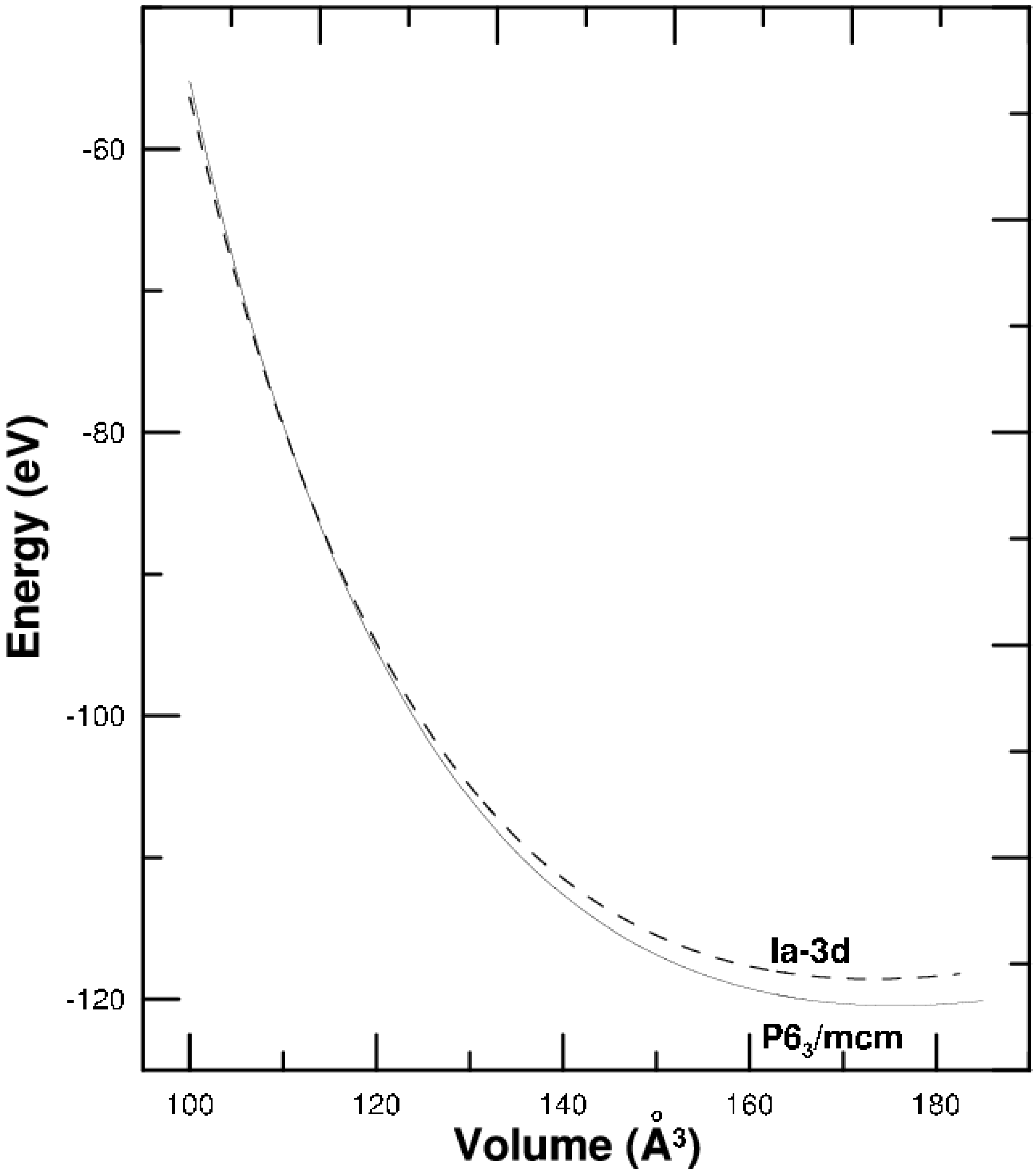

**Figure 7**

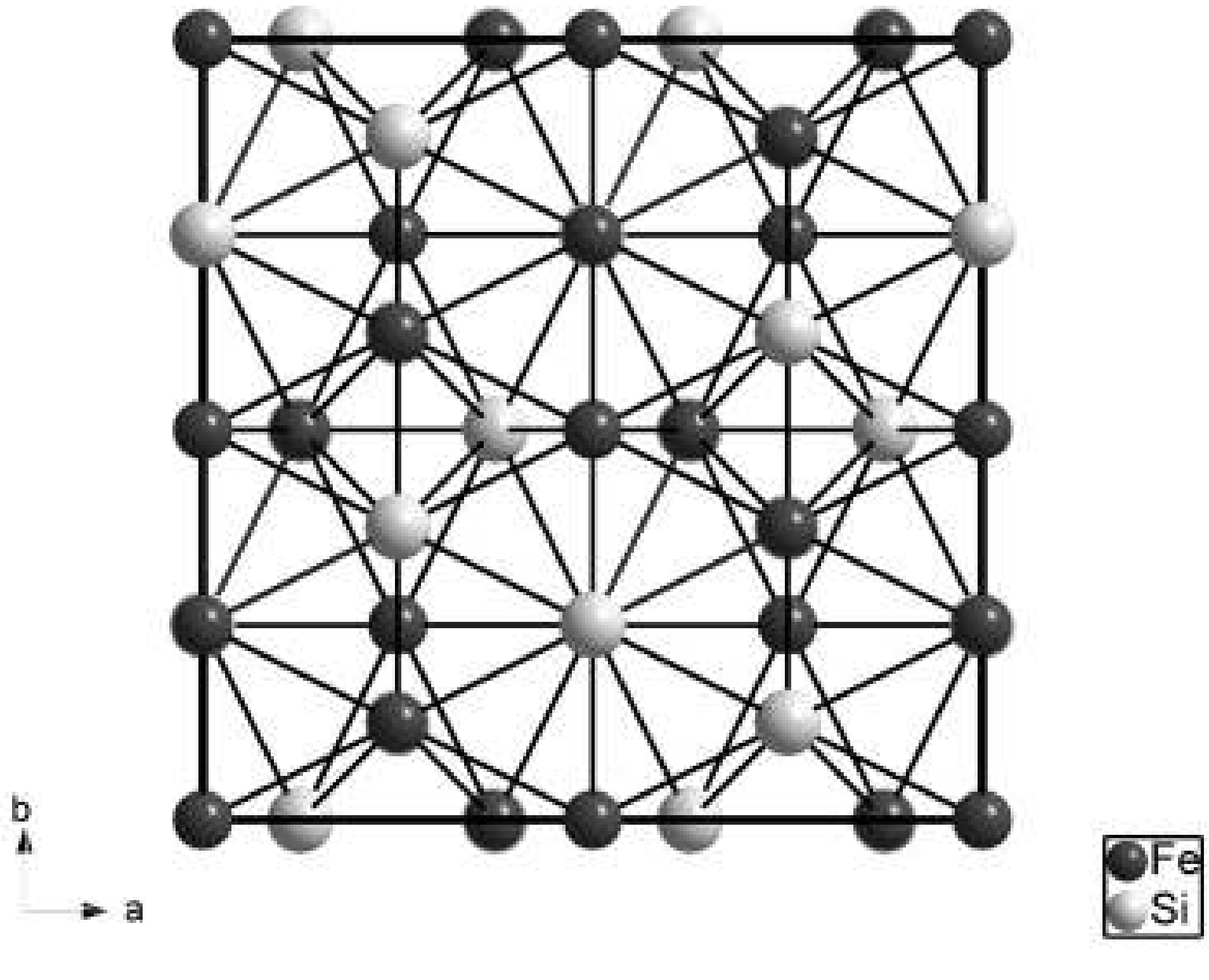

**Figure 8**

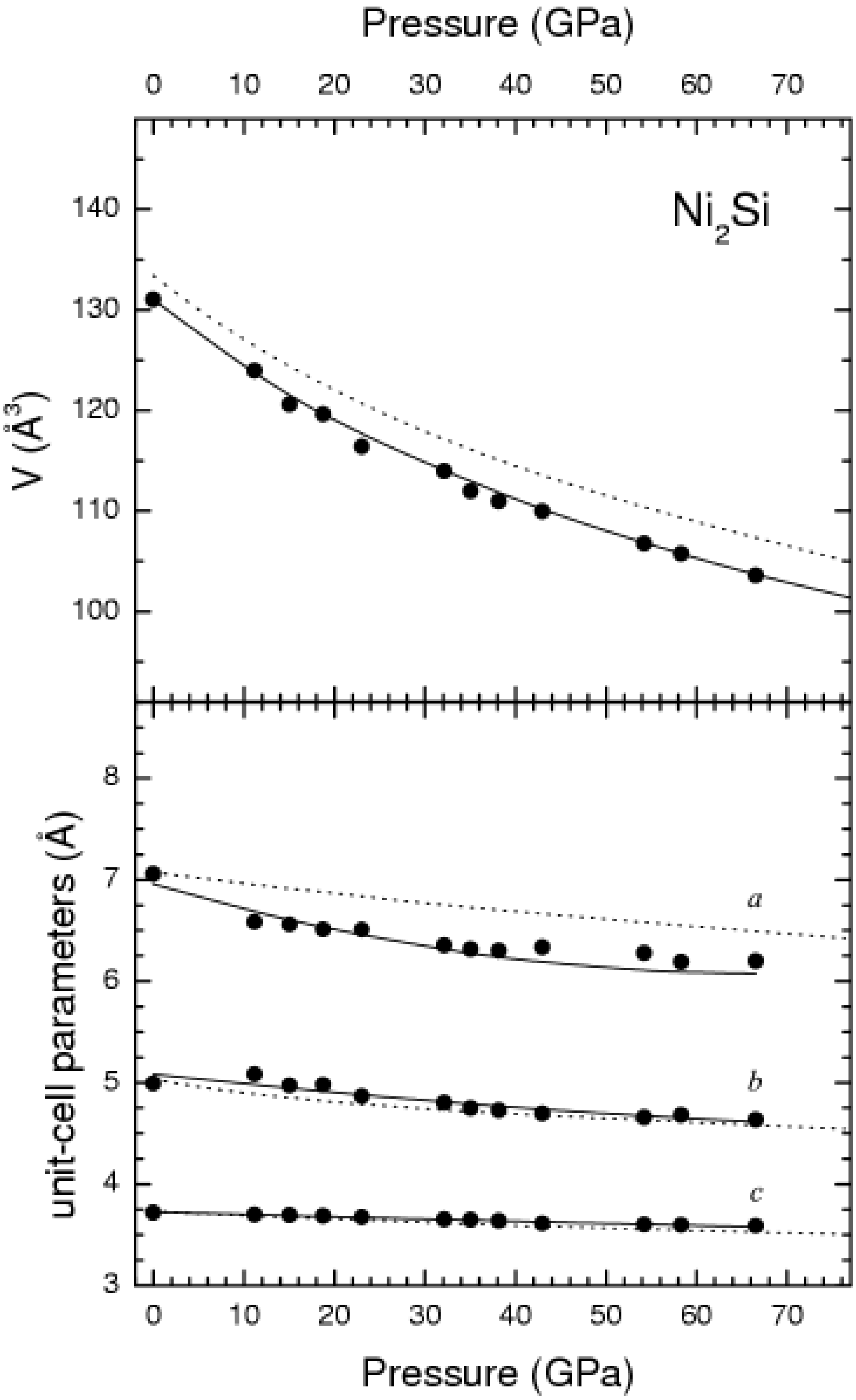

**Figure 9**

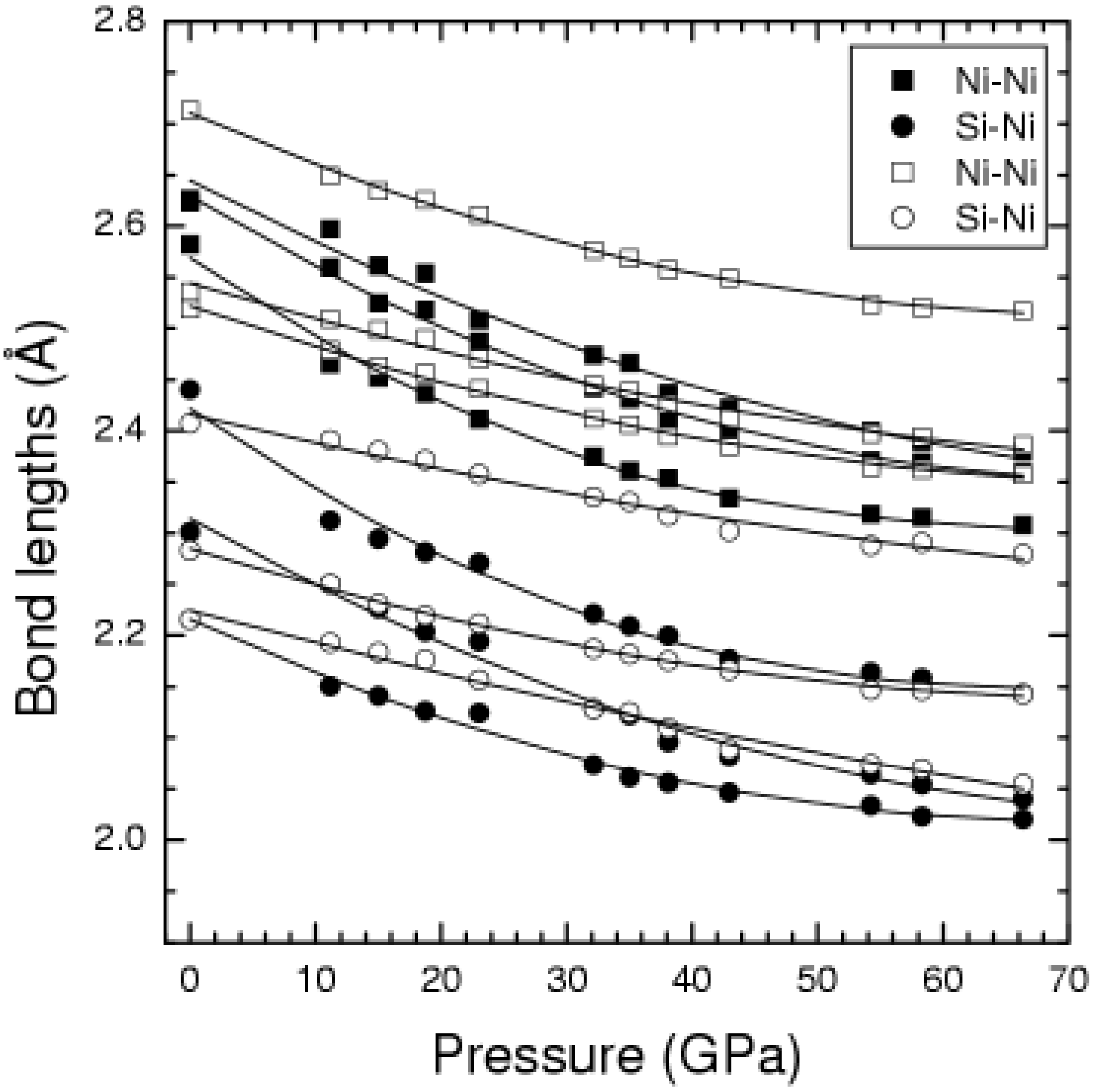

**Figure 10**

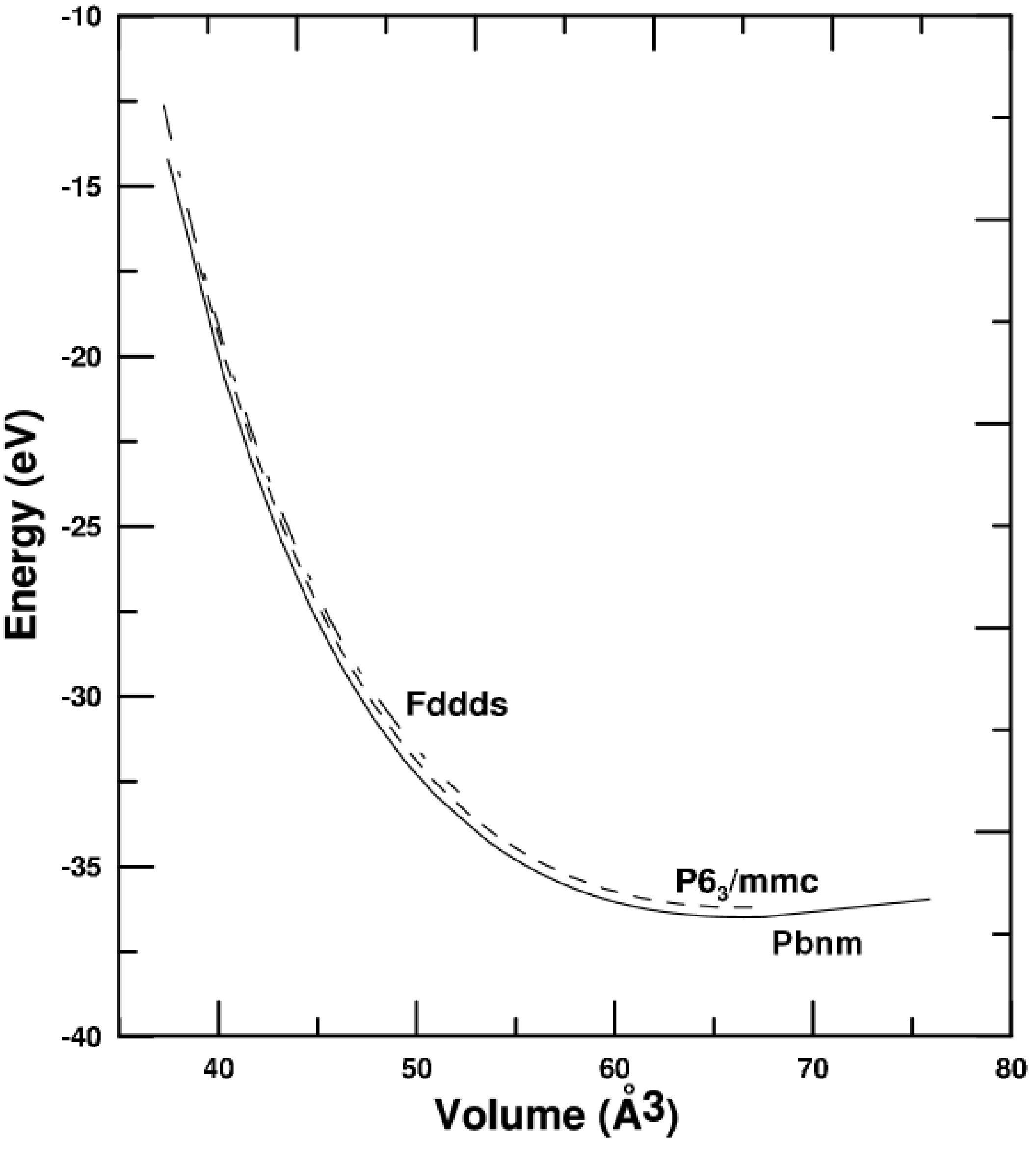

**Figure 11**

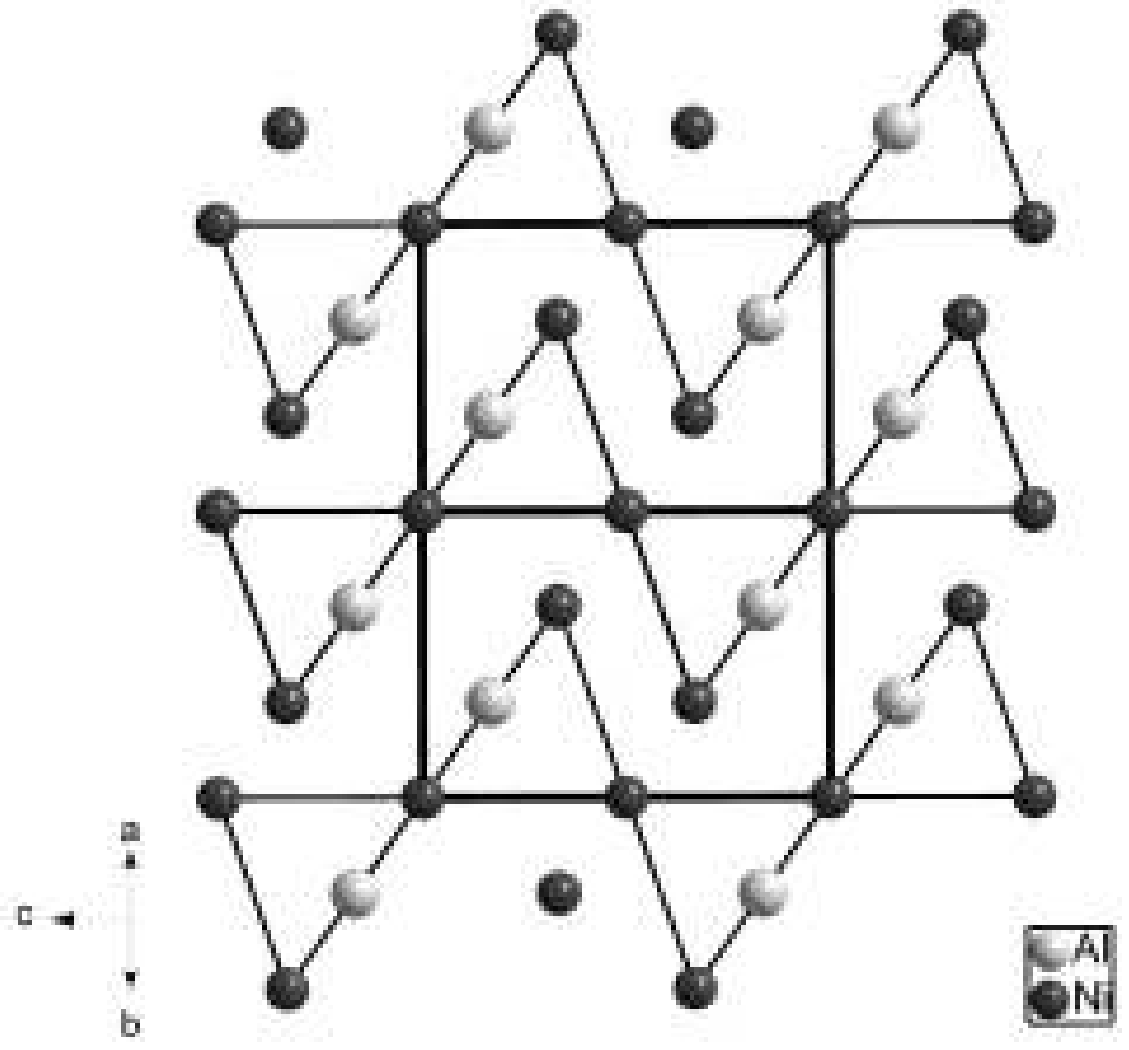

**(a)**

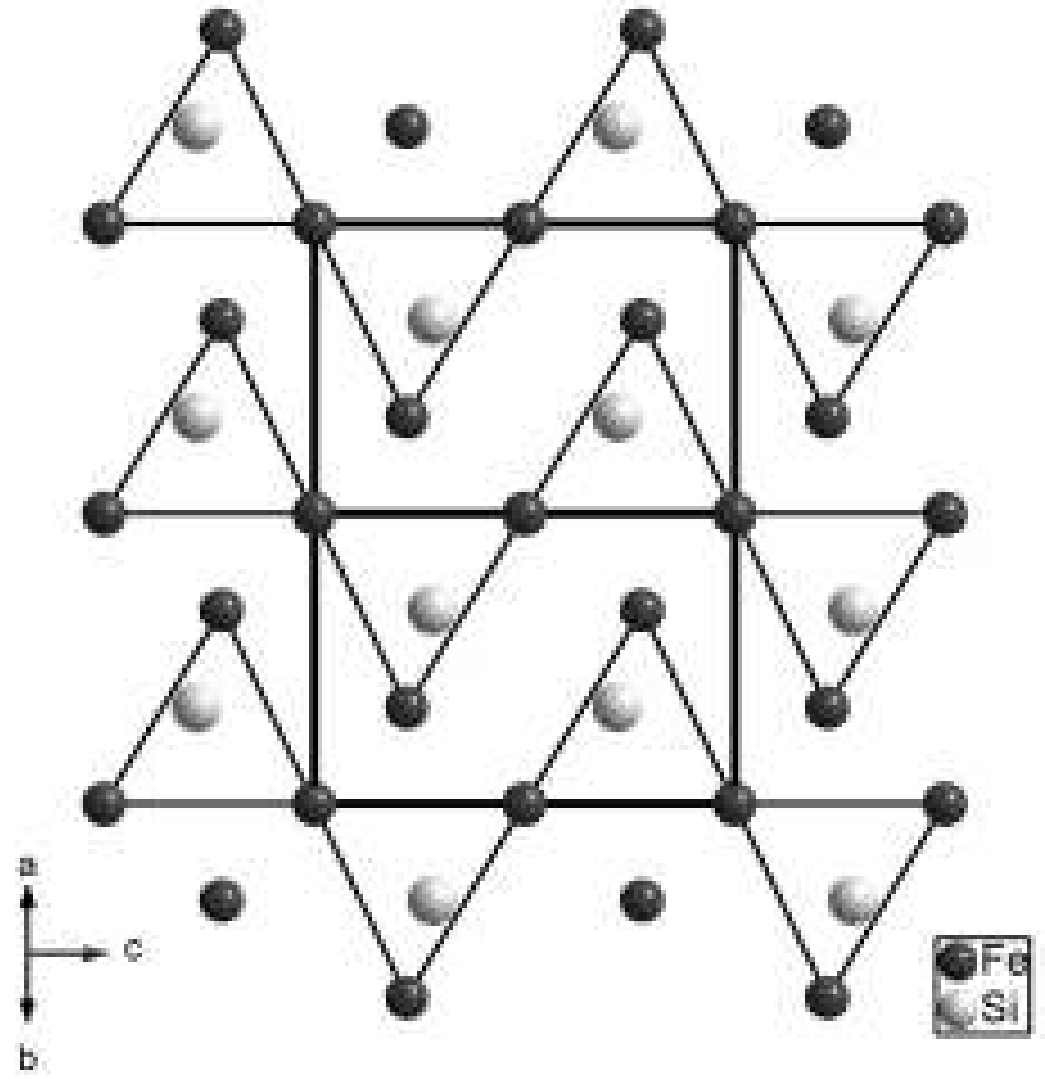

**(b)**

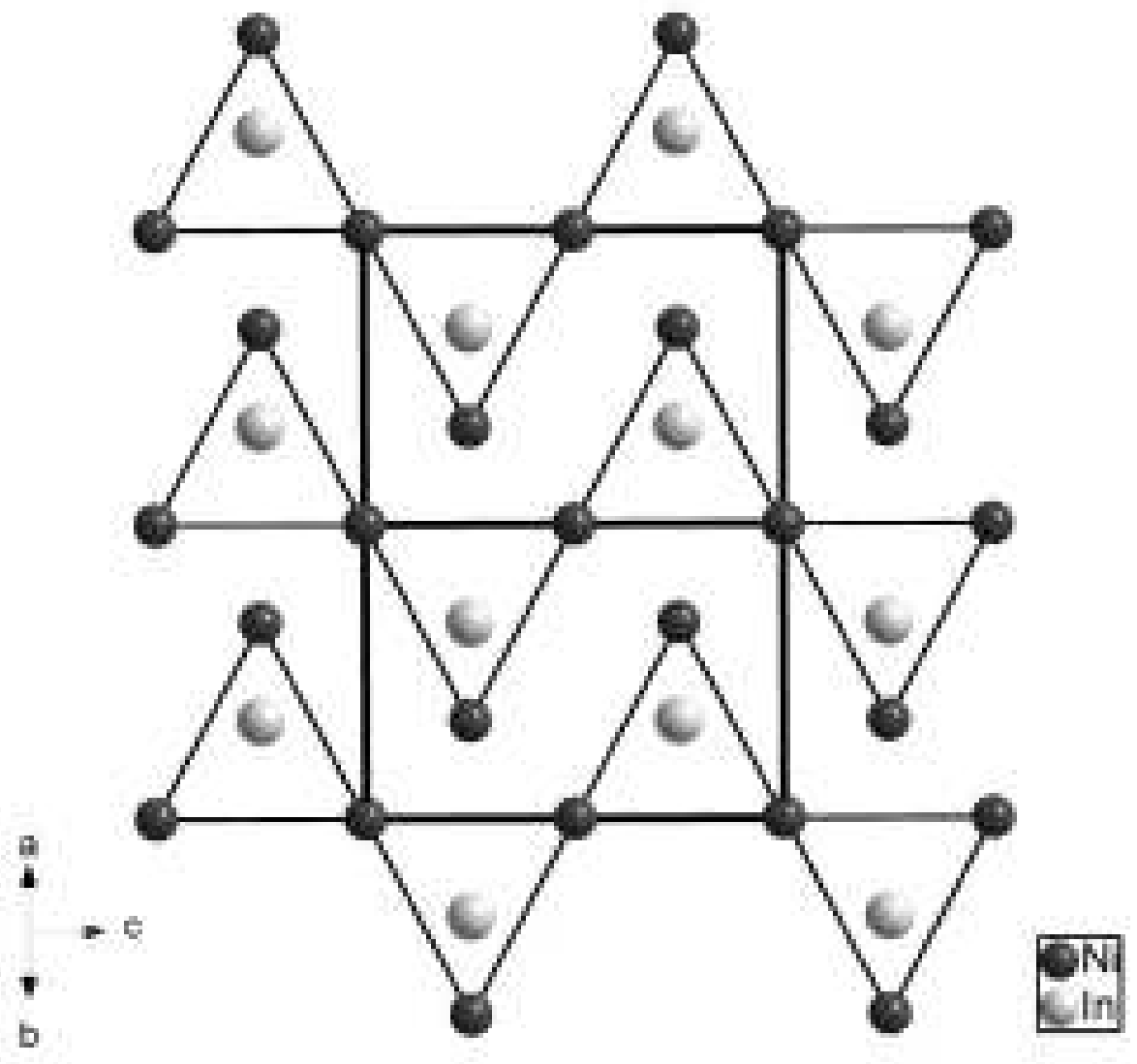

**(c)**

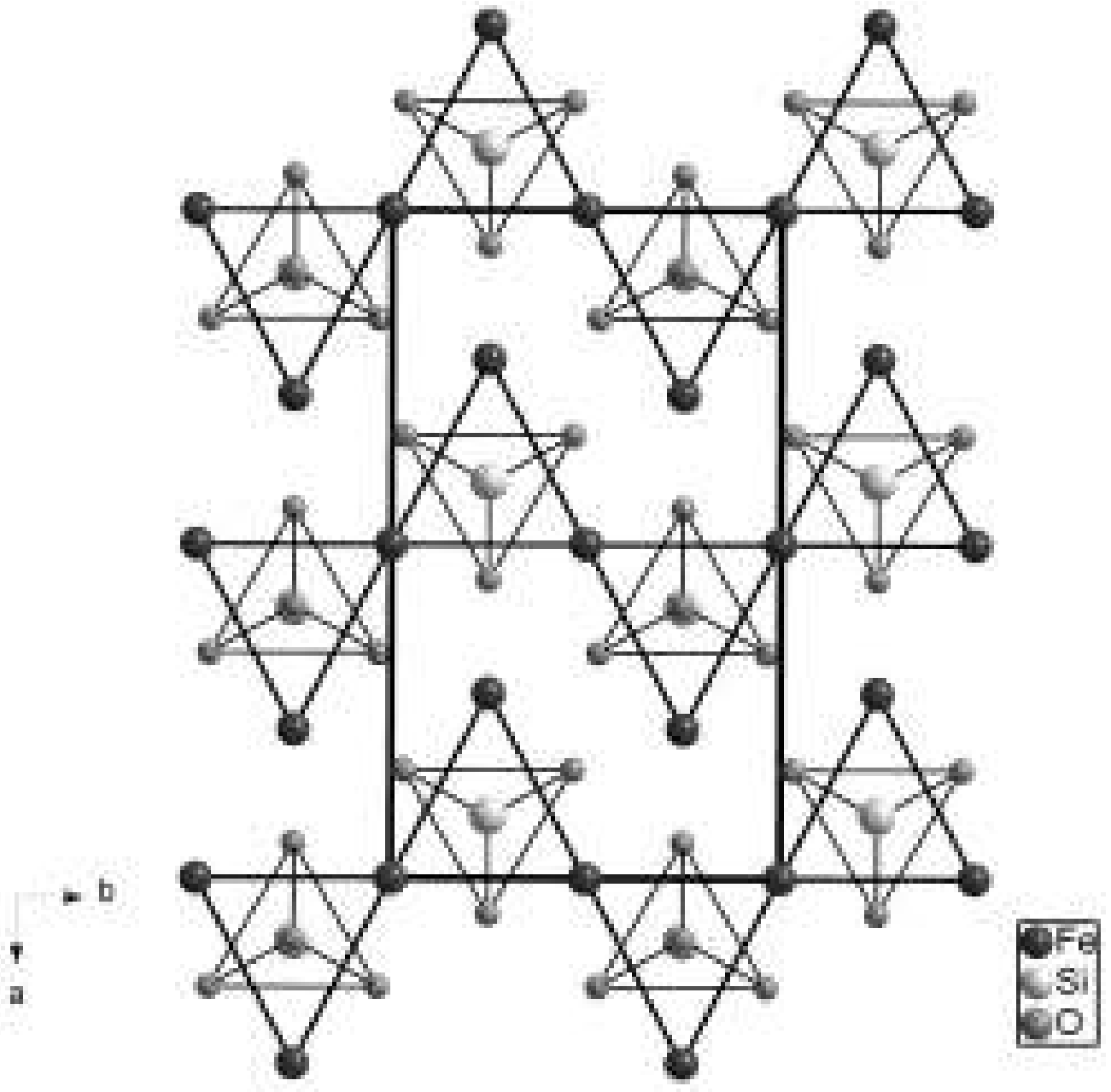

**(d)**

Figure 12

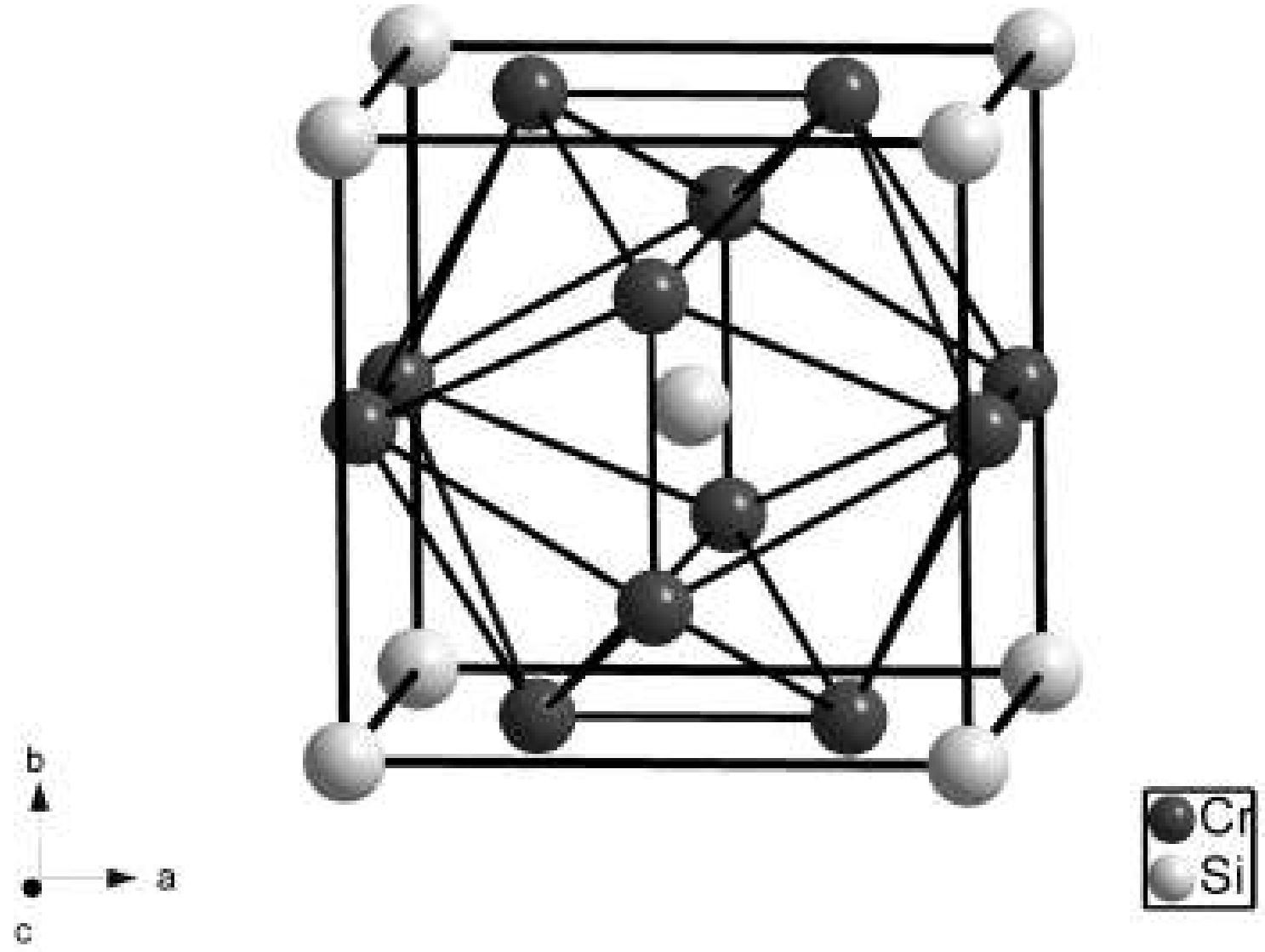

**(a)**

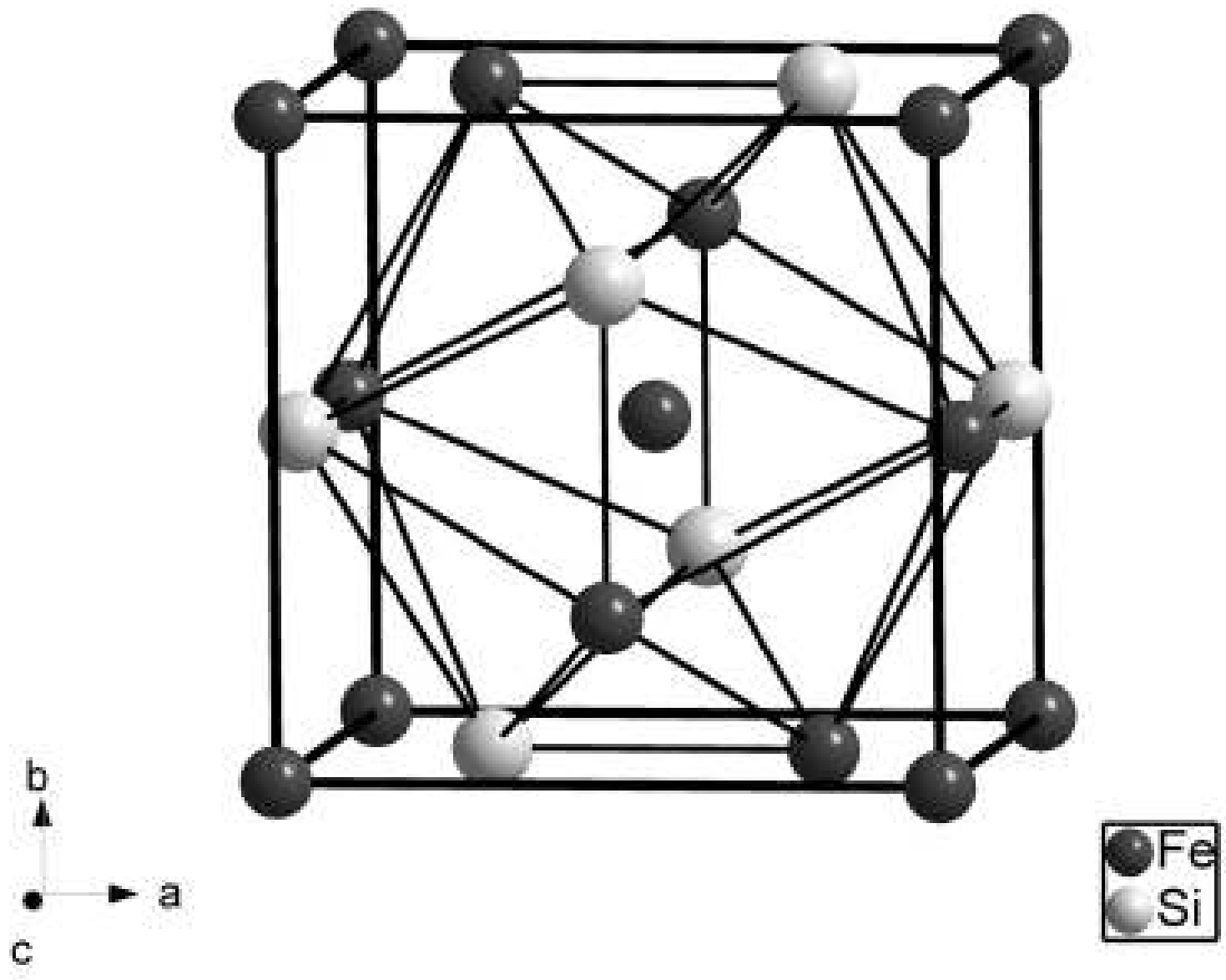

**(b)**

**Figure 13**

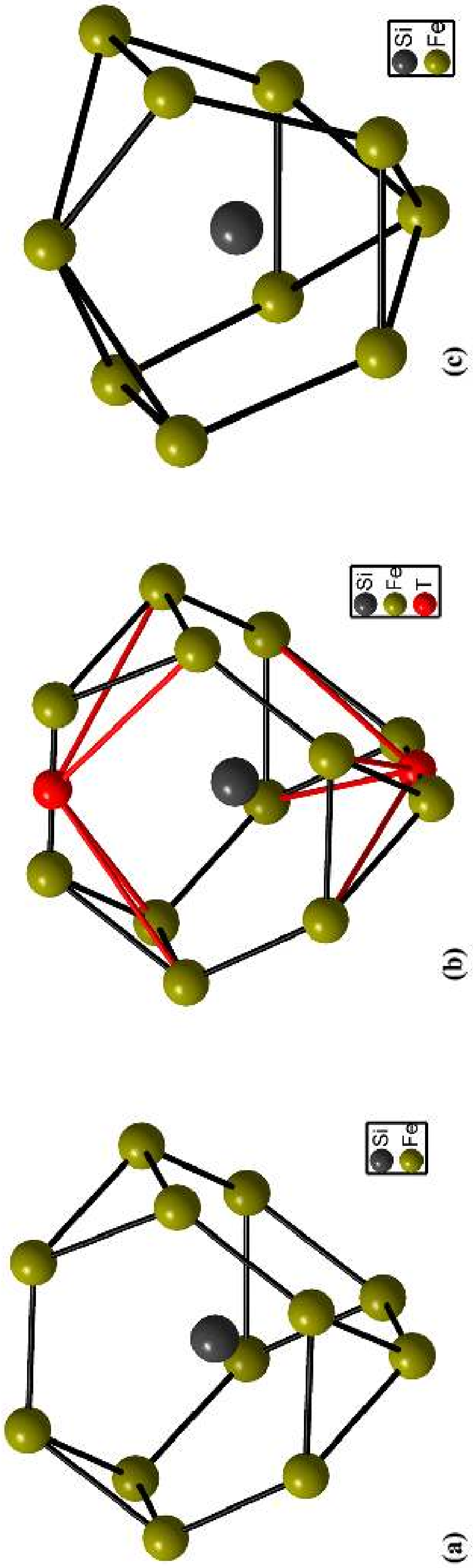